\documentclass[a4paper,11pt]{article}
\usepackage[left=2cm,right=2cm,top=2cm,bottom=2cm]{geometry}
\usepackage[usenames,dvipsnames,svgnames,table]{xcolor}
\usepackage{psfrag,amsmath,amsfonts,amssymb,latexsym,amsthm,verbatim,color,lscape}
\usepackage{relsize}
\usepackage[figuresright]{rotating}
\usepackage[compress]{natbib}
\usepackage{url}
\usepackage{hyperref}
\hypersetup{colorlinks,
citecolor=black,
filecolor=black,
linkcolor=black,
urlcolor=black,
pdftex}

\usepackage{booktabs}
\usepackage[displaymath, mathlines,pagewise]{lineno}

\usepackage{colortbl}
\definecolor{gray}{rgb}{0.9,0.9,0.9}

\usepackage{times}
\usepackage{blindtext}
\usepackage[font={footnotesize,it}]{caption}

\usepackage{sectsty}  
\subsectionfont{\normalsize\bf}
\sectionfont{\large\bf}

\def\mubf{\boldsymbol{\mu}}
\def\Cbar{{\overline C}}

\def\s{\sigma}

\def\de{\delta}
\def\De{\Delta}

\def\Debf{\boldsymbol{\Delta}}

\def\th{\theta}

\long\def\symbolfootnote[#1]#2{\begingroup
\def\thefootnote{\fnsymbol{footnote}}\footnote[#1]{#2}\endgroup}

\DeclareUnicodeCharacter{200E}{}

\begin{document}
\title{The bivariate $K$-finite  normal mixture ``blanket" copula}

\author{
Aristidis K. Nikoloulopoulos\footnote{{\small\texttt{A.Nikoloulopoulos@uea.ac.uk}}, School of Computing Sciences, University of East Anglia, Norwich NR4 7TJ, UK} }
\date{}

\maketitle

\begin{abstract}
\baselineskip=23pt
 \noindent There exist  many bivariate parametric copulas  to model bivariate data  with different dependence features. We propose a new  bivariate parametric copula family that cannot only  handle various dependence patterns that appear in the existing parametric bivariate copula families, but also provides a more enriched  dependence structure.  The proposed copula construction exploits  finite mixtures of bivariate normal distributions. The mixing operation,  the distinct correlation  and mean parameters  at each mixture component  introduce quite a flexible dependence.  The new parametric copula is theoretically investigated, compared  with a set of classical bivariate parametric copulas and  illustrated on two empirical examples from astrophysics and  agriculture 
 where some of the variables have  peculiar   and  asymmetric dependence, respectively.

\end{abstract}

\maketitle

\baselineskip=22pt

\section{Introduction}
Multivariate response data abound in many applications including insurance, risk management, finance, health and environmental sciences. Data from these application areas have different dependence structures including features such as tail dependence \citep{joe93}, that is dependence among extreme values.  Modelling dependence among multivariate outcomes is an interesting problem in statistical science. 
The dependence between random variables is completely described by their multivariate distribution. One may create multivariate distributions based on particular assumptions thus, limiting their use. For example, most existing multivariate distributions assume margins of the same form (e.g., Gaussian, Poisson, etc.) or limited dependence (e.g., tail independence, positive dependence, etc.). 

To solve this problem, copula functions  \citep{joe97,joe2014,nelsen06}  seem to be a promising solution. The power of copulas for dependence modelling is due to the dependence structure being considered separate from the univariate margins. Copulas are a useful way to model multivariate data as they account for the dependence structure and provide a flexible representation of the multivariate distribution. They allow for flexible dependence modelling, different from assuming simple linear correlation structures and normality, which makes them well suited to the aforementioned application areas. In particular, the theory and application of copulas have become important in finance, insurance and other areas, in order to deal with dependence in the joint tails \citep{joeetal10}.

\begin{figure}[!h]

\begin{center}
\begin{tabular}{cc}
\includegraphics[width=0.5\textwidth]{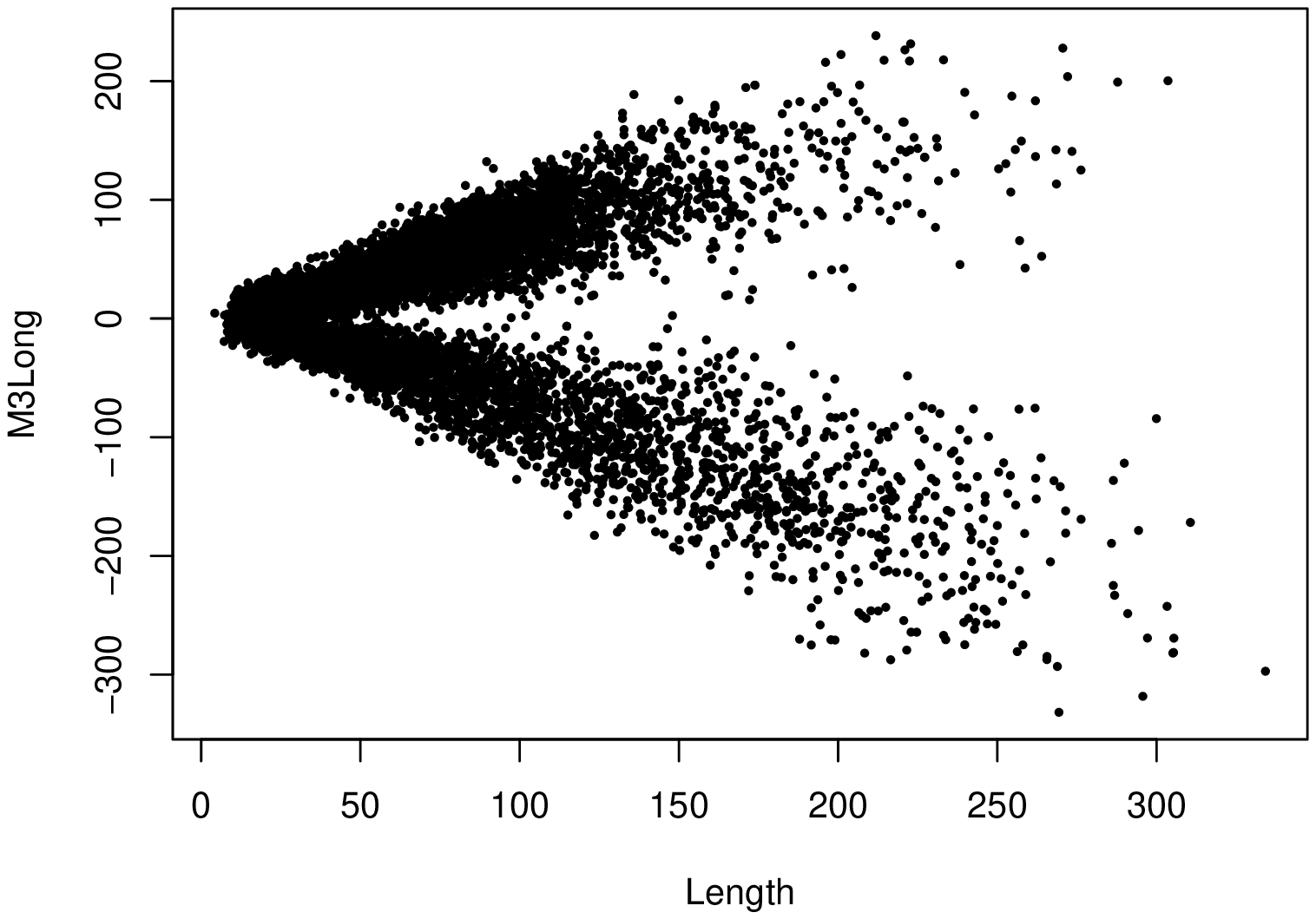}&

\includegraphics[width=0.5\textwidth]{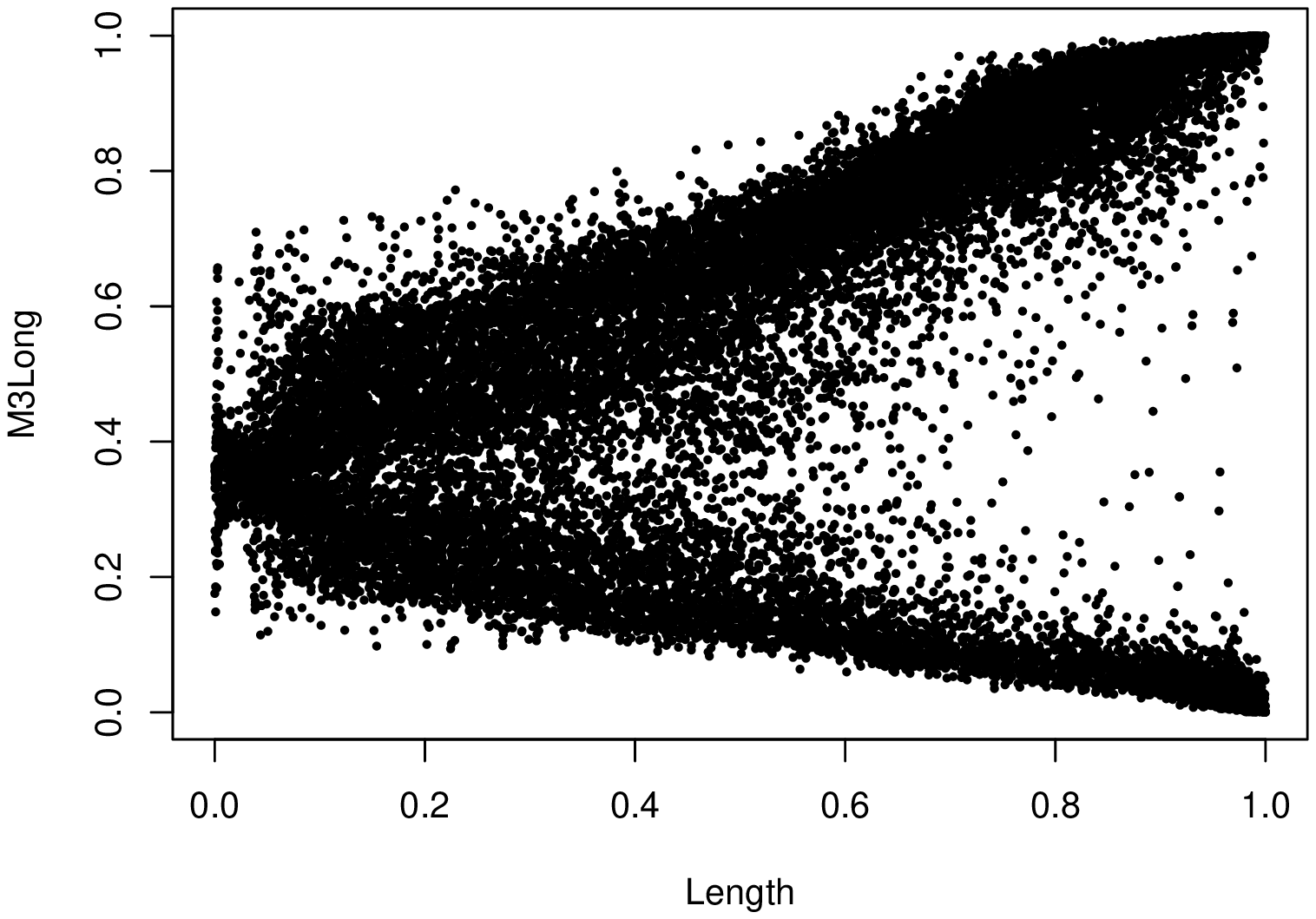} 

\end{tabular}

\caption{\label{data} Scatter plots of the  length of the major axis of the ellipse (Length) and  the third root of the third moment along the major axis (M3Long) on the original and uniform scale.} 

\end{center}
\end{figure}

For the bivariate case, a rich variety of copula families is available and their properties are well-established  \citep{joe97,joe2014,nelsen06}.
Nevertheless, a problem in practical applications is to  identify the most plausible parametric family of bivariate copulas for dependence modelling  \citep{Durrlemanetal2000,
huard&evin&favre05,Nikoloulopoulos&karlis08CSDA}. 
Furthermore,  sometimes none of the bivariate parametric copula families   provides a good  fit. For example, \cite{nagler&czado-2016-jmva}  and \cite{Czado-2019}  analysed the dependence between the Major Atmospheric Gamma Imaging Cherenkov (MAGIC) telescope  variables \citep{bock-etal-2004} and pointed out the uncommon characteristics of the dependence structure between some of the variables, which do not correspond to any bivariate parametric copula. 
 Figure \ref{data} depicts, in particular the relationship  between the length of the major axis of the ellipse (Length) and  the third root of the third moment along the major axis (M3Long), and indeed reveals that none of the existing parametric families of bivariate  copulas (see e.g.,  \citealt{joe97,joe2014}) can model the joint distribution of these variables. Note in passing that in the right panel graph of Figure \ref{data}, we transform the data to the uniform scale by applying their empirical distributions,  in order to isolate the effect of the marginal distributions and solely focus on the dependence structure.

In this paper,  
we propose a new parametric family of copulas that can represent the dependence structure between the aforementioned MAGIC variables. It can also remedy the copula selection issue as can  ``nearly" approximate any parametric family of bivariate copulas. 
A multivariate 2-finite normal mixture (FNM) copula 
has been proposed by \cite{Nikoloulopoulos&karlis07FNM} 
to model multivariate discrete data.   The  correlation matrix for each mixture component was restricted to the identity matrix with the mixing operation introducing the dependence among the discrete responses. Therefore, it has a rather simple computational form, but  suffers from a restricted range of attainable dependence. We will study the  full dependence capacity of the bivariate $K$-FNM copula, where $K$ is the number of mixture components,  by using general correlation matrices for each mixture component and we will show that the the proposed $K$-FNM is a ``blanket" copula, i.e., a copula that can ``nearly" approximate any bivariate parametric copula. 
A similar construction in the literature, which is called Bayesian non-parametric estimation of a copula  \citep{wu-etal-2015-jscs,Vale-etal-2018-jrssc}, takes BVN copulas as the mixture components, hence it allows only for reflection symmetric dependence  
and is not as general as the proposed $K$-FNM copula, which can allow reflection asymmetric dependence. 

The remainder of the paper proceeds as follows. Section \ref{FNMsection}   introduces the bivariate $K$-FNM copula, discusses its properties and provides computational details for maximum likelihood (ML) estimation.  Before that it has a brief overview of
relevant copula theory. 
 Section \ref{blanketSec} shows that the proposed copula is  a ``blanket" copula.   Section \ref{simulations}  studies the small-sample efficiency of the proposed ML estimation technique. 
Section \ref{Appsection} illustrates the  methods  on two empirical examples from astrophysics and  agriculture 
 where some of the variables have  peculiar   and  asymmetric dependence, respectively.  We conclude with some discussion in Section \ref{sec-discussion}.

\section{\label{FNMsection}The bivariate $K$-finite normal mixture copula}
In this section we will define  the bivariate $K$-FNM  copula and study its properties.  Before that, the  first subsection has some background on bivariate copulas. 
\subsection{Overview and relevant background for copulas}
A copula is a multivariate cumulative distribution function (cdf) with uniform $U(0,1)$ margins \citep{joe97,nelsen06,joe2014}.
If $F_{12}$ is a bivariate cdf with univariate margins $F_1,F_2$,
then Sklar's (1959) \nocite{sklar1959} theorem implies that there is a copula $C$ such that
\begin{equation*}\label{copulacdf}
F_{12}(y_1,y_2)= C\Bigl(F_1(y_1),F_2(y_2)\Bigr).
\end{equation*}
The copula is unique if $F_1,F_2$ are continuous, but not
if some of the $F_j$ have discrete components.
If $F_{12}$ is continuous and $(Y_1,Y_2)\sim F_{12}$, then the unique copula
is the distribution of $(U_1,U_2)=\left(F_1(Y_1),F_2(Y_2)\right)$ leading to
  \begin{equation}\label{InverseSklar}
 C(u_1,u_2)=F_{12}\Bigl(F_1^{-1}(u_1),F_2^{-1}(u_2)\Bigr),
  \quad 0\le u_j\le 1, j=1,2, \end{equation}
where $F_j^{-1}$ are inverse cdfs. In particular, if $\Phi_{12}(\cdot;\th)$
is the bivariate normal (BVN) cdf with correlation $\th$ and
standard normal margins, and $\Phi$ is the univariate standard normal cdf,
then the BVN copula is
 \begin{equation}\label{BVNcopula-cdf}
C(u_1,u_2)=\Phi_{12}\Bigl(\Phi^{-1}(u_1),\Phi^{-1}(u_2);\th\Bigr).
\end{equation}

If $C(\cdot;\theta)$ is a parametric
family of copulas and $F_j(\cdot;\eta_j)$ is a parametric model for the
$j$th univariate margin, then
  $$C\Bigl(F_1(y_1;\eta_1),F_2(y_2;\eta_2);\theta\Bigr)$$
is a bivariate parametric model with univariate margins $F_1,F_2$.
For copula models, the variables can be continuous or discrete  \citep{Nikoloulopoulos2013a,nikoloulopoulos&joe12}.

\subsection{The bivariate $K$-FNM copula}

Let a bivariate $K$-FNM distribution be defined as
\begin{equation*}\label{mfinite}
\sum_{k=1}^{K}\pi_k {\mathcal N}_2 (
\mubf_k,\mathbf{\Sigma}_k), \quad 0<\pi_k < 1,\quad
\sum_{k=1}^{K}\pi_k=1,
\end{equation*}
where ${\mathcal N}_2(\mubf,\mathbf{\Sigma})$ denotes
the BVN
distribution with mean vector $\mubf=(\mu_1,\mu_2)$ and covariance matrix
$\mathbf{\Sigma} =\begin{pmatrix}\s_1^2& \rho\s_1\s_2\\
 \rho\s_1\s_2&\s_2^2\end{pmatrix}$.
Its cdf  is given
by
\begin{equation}\label{FNMcdf}
\mathcal{F}_2(\mathbf{y};\pi_k,\mubf_k,\mathbf{\Sigma}_k,\,k=1,\ldots,K) =
\sum_{k=1}^{K}\pi_k \Phi_2(\mathbf{y};
\mubf_k,\mathbf{\Sigma}_k), \quad 0<\pi_k < 1,\quad
\sum_{k=1}^{K}\pi_k=1,
\end{equation}
where $\Phi_2(\mathbf{y}; \mubf,\mathbf{\Sigma})$ is the cdf
of the ${\mathcal N}_2 (\mubf, \mathbf{\Sigma})$
distribution.

From (\ref{InverseSklar}) if $F_{12}$ is the bivariate $K$-FNM cdf $\mathcal{F}_2$ in (\ref{FNMcdf}) and $F_1=F_2=\mathcal{F}$, where $\mathcal{F}$ is the univariate $K$-FNM  cdf,
then the bivariate $K$-FNM copula is defined as 
\begin{multline}
C(u_1,u_2;\pi_k,\mubf_k,\mathbf{\Sigma}_k,\,k=1,\ldots,K)=\\
\mathcal{F}_2\Bigl(\mathcal{F}^{-1}(u_1;\pi_k,\mu_{k1},\s_{k1},\,k=1,\ldots,K),\mathcal{F}^{-1}(u_2;\pi_k,\mu_{k2},\s_{k2},\,k=1,\ldots,K);\pi_k,\mubf_k,\mathbf{\Sigma}_k,\,k=1,\ldots,K\Bigr).
\end{multline}
Subsequently, one can derive the bivariate $K$-FNM copula density as below
\begin{multline}\label{FNMdensity}
c(u_1,u_2;\pi_k,\mubf_k,\mathbf{\Sigma}_k,\,k=1,\ldots,K)=\\
\frac{{f}_2\Bigl(\mathcal{F}^{-1}(u_1;\pi_k,\mu_{k1},\s_{k1},\,k=1,\ldots,K),\mathcal{F}^{-1}(u_2;\pi_k,\mu_{k2},\s_{k2},\,k=1,\ldots,K);\pi_k,\mubf_k,\mathbf{\Sigma}_k,\,k=1,\ldots,K\Bigr)}{f\Bigl(\mathcal{F}^{-1}(u_1;\pi_k,\mu_{k1},\s_{k1},\,k=1,\ldots,K)\Bigr)f\Bigl(\mathcal{F}^{-1}(u_2;\pi_k,\mu_{k2},\s_{k2},\,k=1,\ldots,K)\Bigr)},
\end{multline}
where $f$ and $f_2$ is the univariate and bivariate density, respectively, of the  $K$-FNM distribution.

\subsection{Dependence properties of the $K$-FNM distribution}
We  study the dependence properties of the bivariate $K$-FNM distribution as these will be inherited to the copula. The mean vector and covariance matrix of the $K$-FNM are given respectively by
 $$
\mubf=\sum_{k=1}^{K}\pi_k\mubf_k \quad \mbox{and}
\quad \Debf=\sum_{k=1}^{K}\pi_k\mathbf{\Sigma}_k +
\sum_{k=1}^{K}\pi_k\mubf_k\mubf_k^\top
-\mubf\mubf^\top.
 $$ 
An aspect of mixture models is their  lack of identifiability, but this can be overcome by imposing some restrictions in the parameters. 
In our approach, to overcome the typical identifiability issues we priory assume that $\mubf_1+\ldots+\mubf_K={\bf 0}$, i.e., the mean vectors become 
$$\mubf_1=(\nu_1,\nu_2),  \ldots, \mubf_{K-1}=(\nu_{2K-3},\nu_{2K-2}),
\mubf_{K}=(-\sum_{k=1}^{K-1}\nu_{2k-1},-\sum_{k=1}^{K-1}\nu_{2k}),$$ and that the variances of the mixture components are set to one, i.e., $\s_{k1}^2=\s_{k2}^2=1$ for $k=1,\ldots,K$.

The covariance matrix is then of the form $
\Debf=\begin{pmatrix}
\De_{11} & \De_{12}\\
\De_{12} & \De_{22}
\end{pmatrix},
$
where 
\begin{equation*}
\De_{11}=1+\sum_{k=1}^{K-1}\pi_k\nu_{2k-1}^2+\pi_K\Bigl(\sum_{k=1}^{K-1}\nu_{2k-1}\Bigr)^2-\Bigl(\sum_{k=1}^{K-1}\pi_k\nu_{2k-1}-\pi_K\sum_{k=1}^{K-1}\nu_{2k-1}\Bigr)^2,
\end{equation*}

\begin{small}

\begin{equation*}
\De_{12}=\sum_{k=1}^{K-1}\pi_k\nu_{2k-1}\nu_{2k}+\pi_K\sum_{k=1}^{K-1}\nu_{2k-1}\sum_{k=1}^{K-1}\nu_{2k}-
\Bigl(\sum_{k=1}^{K-1}\pi_k\nu_{2k-1}-\pi_K\sum_{k=1}^{K-1}\nu_{2k-1}\Bigr)\Bigl(\sum_{k=1}^{K-1}\pi_k\nu_{2k}-\pi_K\sum_{k=1}^{K-1}\nu_{2k}\Bigr)+\sum_{k=1}^K\pi_k\rho_k,
\end{equation*}
\end{small}

\noindent and 
\begin{equation*}
\De_{22}=1+\sum_{k=1}^{K-1}\pi_k\nu_{2k}^2+\pi_K\Bigl(\sum_{k=1}^{K-1}\nu_{2k}\Bigr)^2-\Bigl(\sum_{k=1}^{K-1}\pi_k\nu_{2k}-\pi_K\sum_{k=1}^{K-1}\nu_{2k}\Bigr)^2.
\end{equation*}

As one can easily see, an identifiability problem still occurs. To overcome this, we set $\nu_1=K-1, \nu_3=\cdots=\nu_{2K-3}=-1$ and $\nu_2=\th_1,\nu_4=\th_2,\ldots,\nu_{2K-2}=\th_{K-1}$. 

Accordingly, the variance-covariance terms of $\Debf$ reduce to 
\begin{equation*}
\De_{11}=
1+\pi_1(1-\pi_1)K^2, 
\end{equation*}

\begin{equation*}
\De_{12}=\pi_1\th_1(K-1)-\sum_{k=2}^{K-1}\pi_k\th_k+\pi_K\sum_{k=1}^{K-1}\th_k+(1-\pi_1K)
\Bigl(\sum_{k=1}^{K-1}\pi_k\th_k-\pi_K\sum_{k=1}^K\th_k\Bigr)+\sum_{k=1}^{K}\pi_k\rho_k,
\end{equation*}
and
\begin{equation*}
\De_{22}=1+\sum_{k=1}^{K-1}\pi_k\th_k^2+\pi_K\Bigl(\sum_{k=1}^{K-1}\th_k\Bigr)^2-\Bigl(\sum_{k=1}^{K-1}\pi_k\th_k-\pi_K\sum_{k=1}^{K-1}\th_k\Bigr)^2.
\end{equation*}

The Pearson's correlation parameter is 
\begin{equation*}
\rho= \frac{\pi_1\th_1(K-1)-\sum_{k=2}^{K-1}\pi_k\th_k+\pi_K\sum_{k=1}^{K-1}\th_k+(1-\pi_1K)
\Bigl(\sum_{k=1}^{K-1}\pi_k\th_k-\pi_K\sum_{k=1}^K\th_k\Bigr)+\sum_{k=1}^{K}\pi_k\rho_k}{\sqrt{1+\pi_1(1-\pi_1)K^2}\sqrt{1+\sum_{k=1}^{K-1}\pi_k\th_k^2+\pi_K\Bigl(\sum_{k=1}^{K-1}\th_k\Bigr)^2-\Bigl(\sum_{k=1}^{K-1}\pi_k\th_k-\sum_{k=1}^{K-1}\th_k\Bigr)^2}}
\end{equation*}
and can attain the $\pm1$ values.

\begin{figure}[!h]
\begin{center}
\begin{tabular}{ccc}\hline
\multicolumn{3}{c}{$K=2 \quad \pi_1=0.3 \quad \rho_1=0.8 \quad \rho_2=-0.8$}\\
$\th_1=-1.5$&$\th_1=0$&$\th_1=1.5$\\\hline
\includegraphics[width=0.3\textwidth]{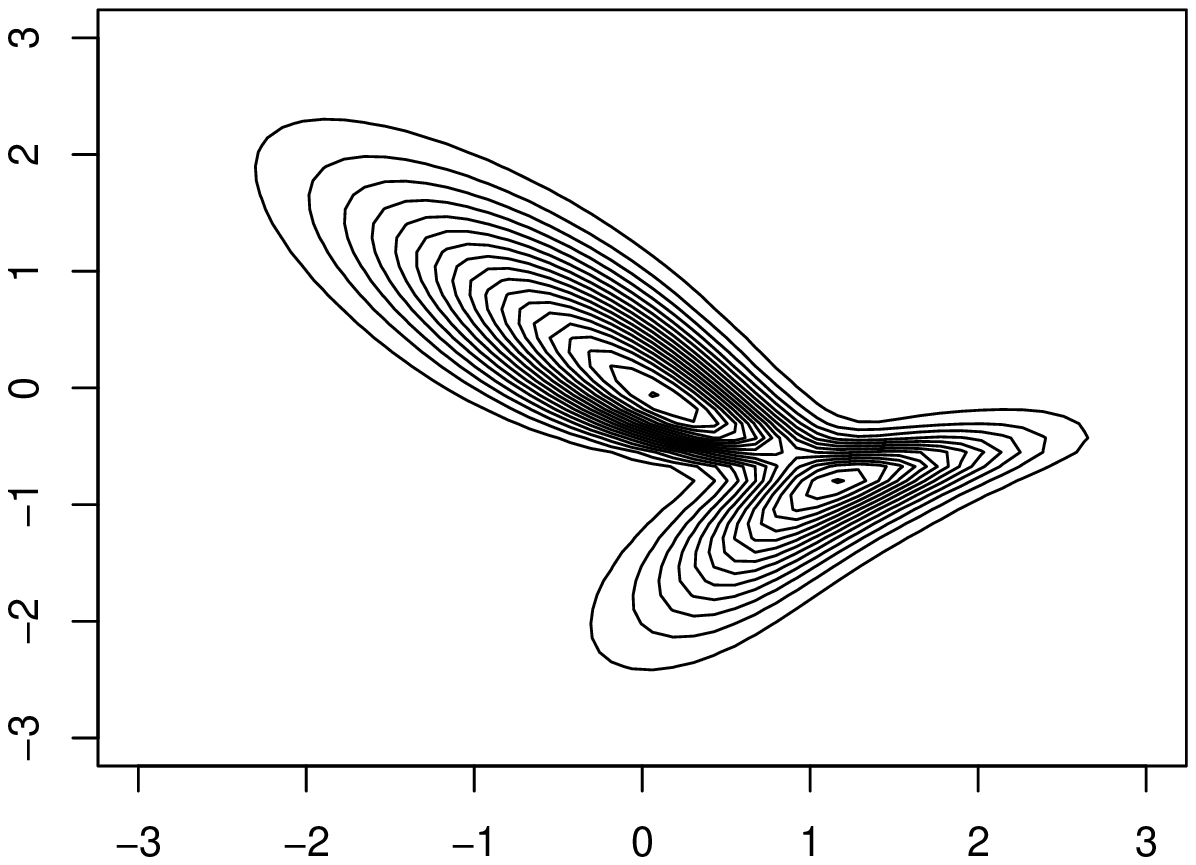}
&
\includegraphics[width=0.3\textwidth]{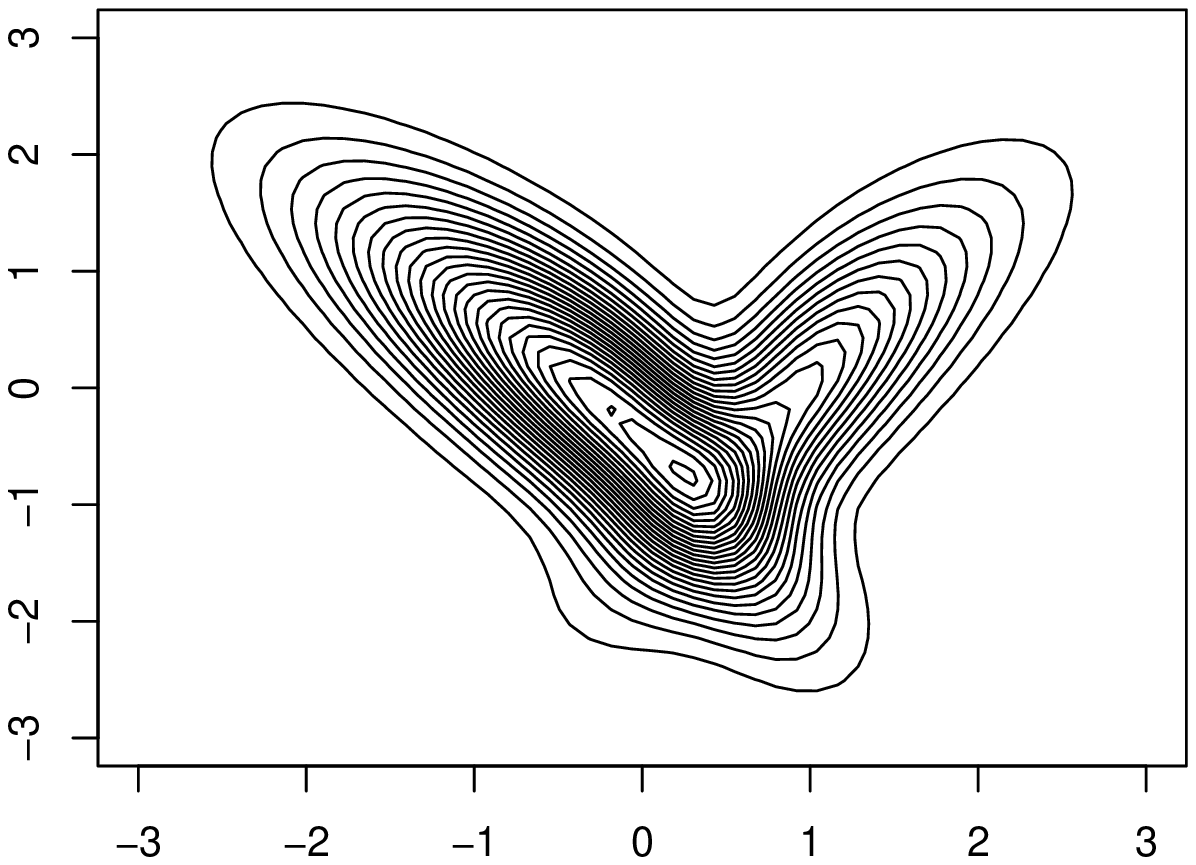}&
\includegraphics[width=0.3\textwidth]{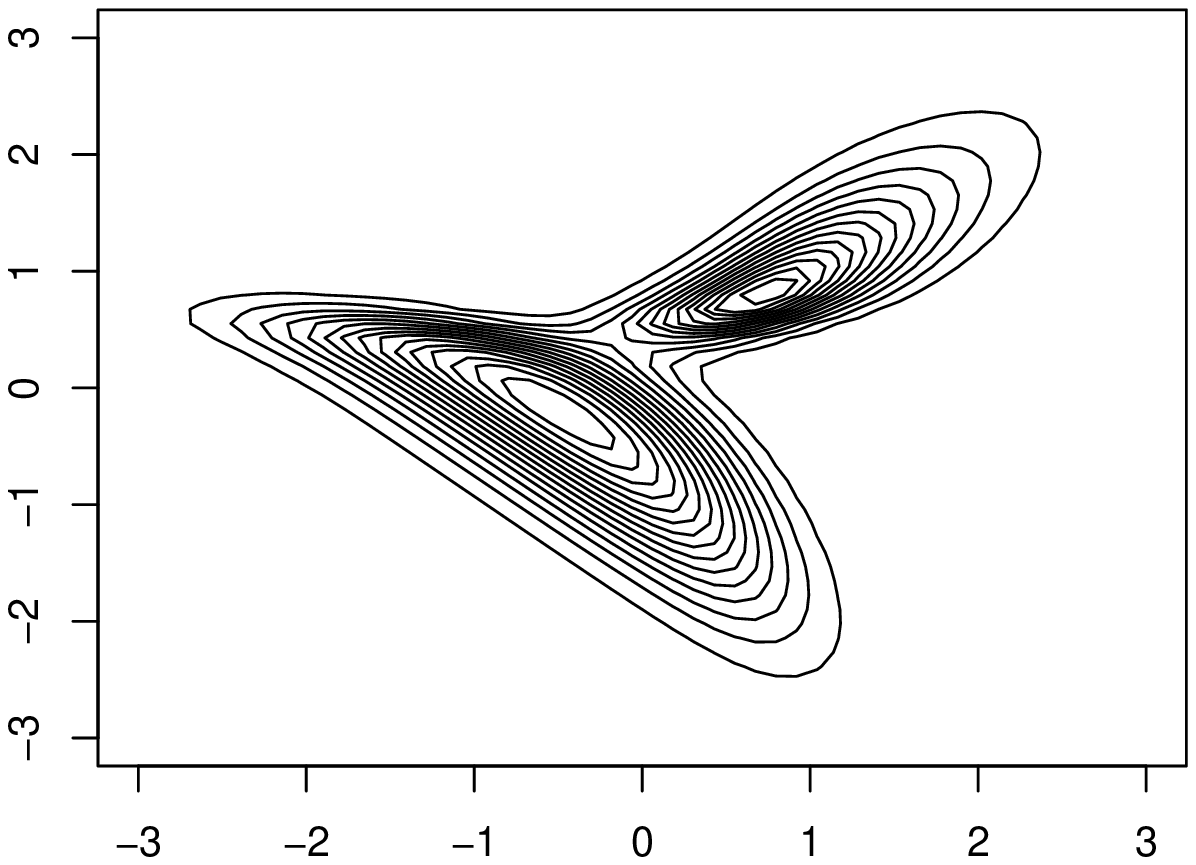}
\\
\hline
\multicolumn{3}{c}{$K=3\quad  \rho_1=0.8 \quad \rho_2=-0.8 \quad \rho_3=0.8 \quad \pi_1=0.2 \quad \pi_2=0.5$}\\
$\th_1=\th_2=-1.5$&$\th_1=\th_2=0$&$\th_1=\th_2=1.5$\\\hline
\includegraphics[width=0.3\textwidth]{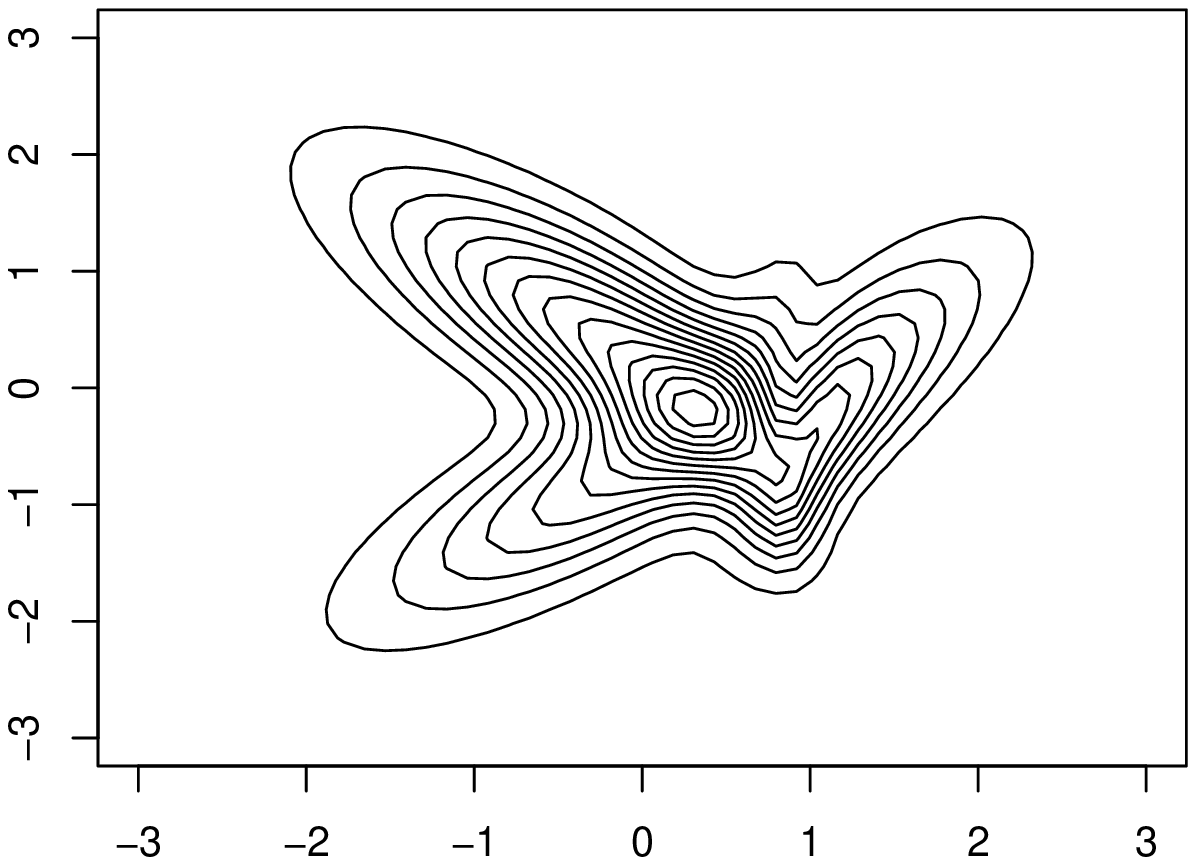}&
\includegraphics[width=0.3\textwidth]{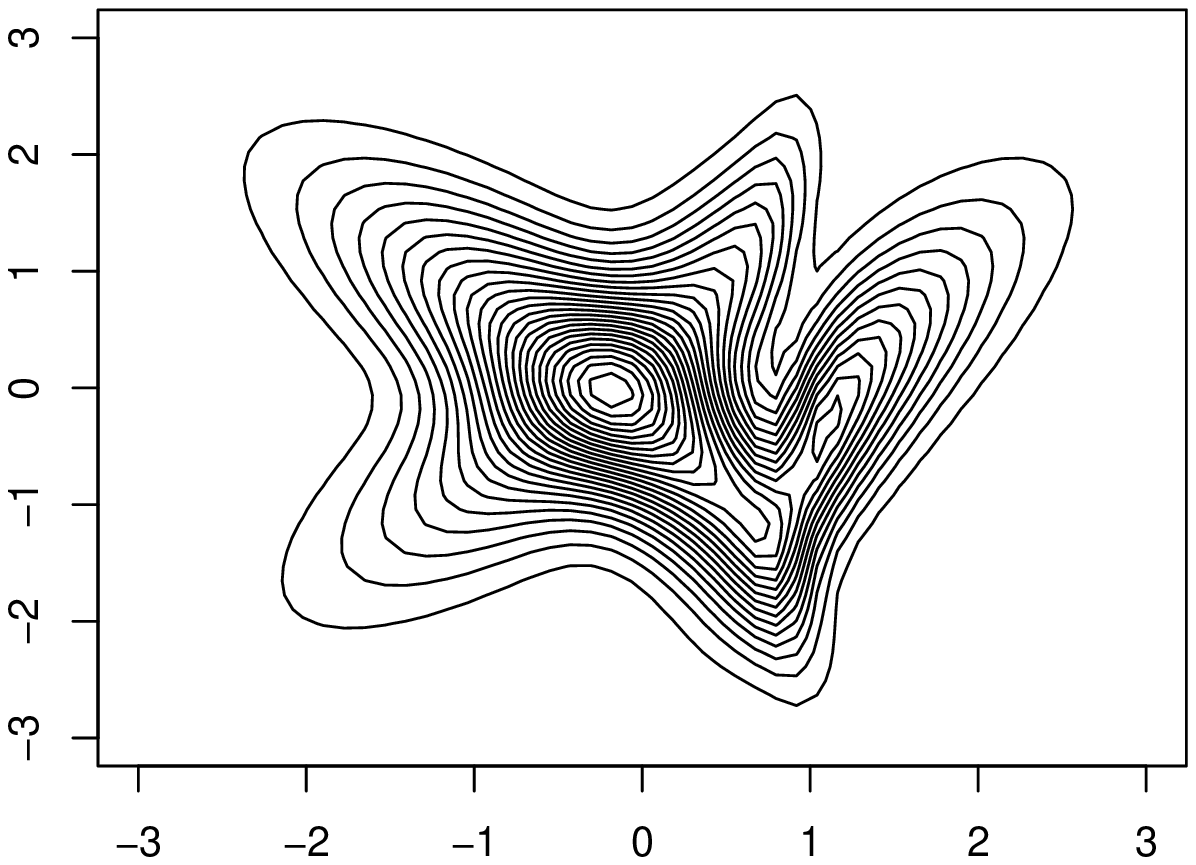}
&
\includegraphics[width=0.3\textwidth]{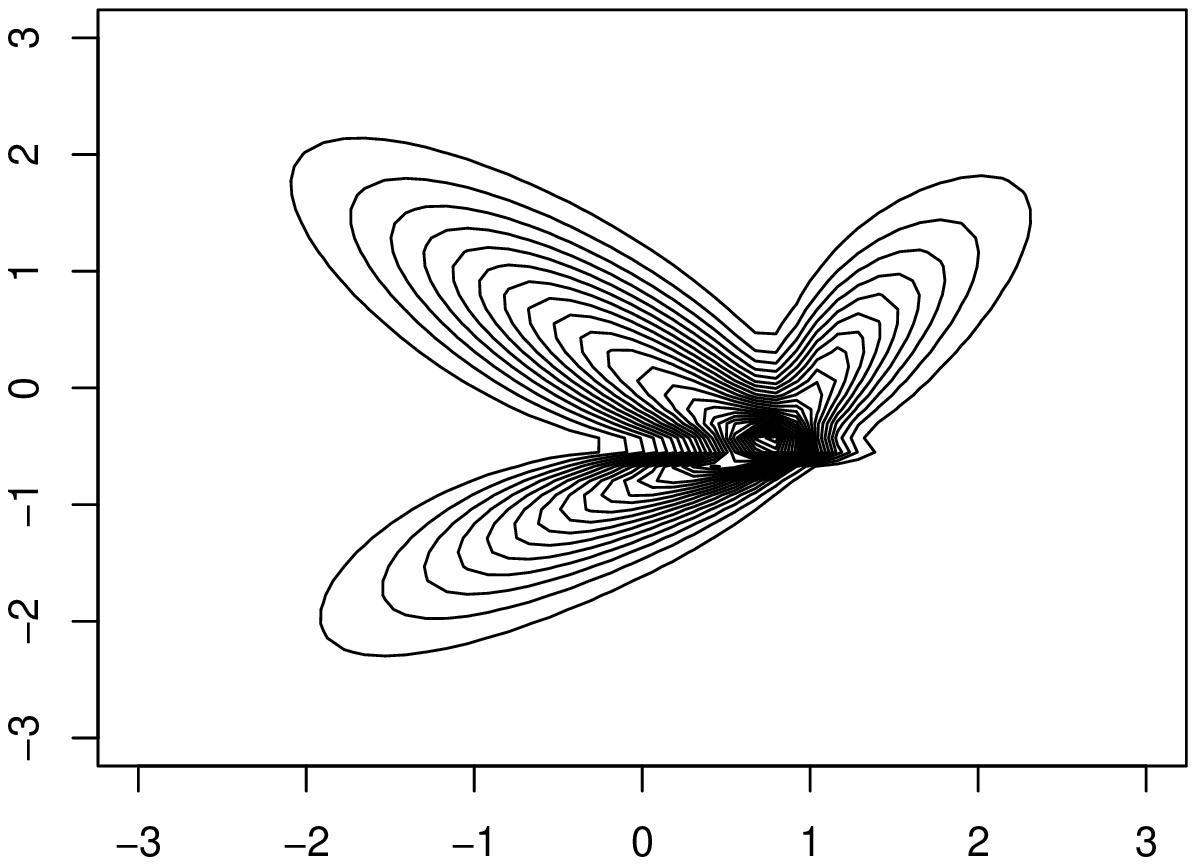}\\\hline
\end{tabular}

\caption{\label{strange-contours}Contour plots of the $K$-FNM copula with $K=2$  (upper panel) and $K=3$ (lower panel) components and standard normal margins
for various
parametrizations.}
\end{center}
\end{figure}

We depict  some dependence shapes that can be  imposed by the bivariate $K$-FNM copula with the above parametrization in Figure \ref{strange-contours}. 

\subsection{\label{MLsection}Maximum likelihood estimation}
In copula models, a copula  is combined with a set of univariate margins. This is equivalent to assuming that variables $Y_1,Y_2$ have been transformed to uniform random variables $U_1=F_1(Y_1),U_2=F_2(Y_2)$.  
For data $y_{ij},\,i=1,\ldots,n,\,j=1,2$, we  use either  non-parametric or parametric univariate distributions to transform the data $y_{ij}$ to copula data $u_{ij}=F_j(y_{ij})$, i.e., data on the uniform scale. 
These semi-parametric and parametric estimation techniques have been developed by \cite{gen&ghoudi&riv95} and \cite{joe05}, respectively,  and  can be  regarded as two-step approaches on the original data or simply as the standard one-step maximum likelihood (ML) method on the transformed (copula) data.

To this end, estimation of the $K$-FNM copula parameters $(\pi_1,\ldots,\pi_{K-1},\th_1,\ldots,\th_{K-1},\rho_1,\ldots,\rho_K)$  can be approached 
by maximizing the logarithm of the joint likelihood 
$$
\ell(\pi_1,\ldots,\pi_{K-1},\th_1,\ldots,\th_{K-1},\rho_1,\ldots,\rho_K)=\sum_{i=1}^N\log\Bigl (c(u_{i1},u_{i2};\pi_1,\ldots,\pi_{K-1},\th_1,\ldots,\th_{K-1},\rho_1,\ldots,\rho_K)\Bigr),
$$
where $c(\cdot;\cdot)$ is the $K$-FNM copula density given in (\ref{FNMdensity}). 
The estimated parameters can be obtained by 
using a quasi-Newton \citep{nash90} method applied to the logarithm of the joint likelihood.  
This numerical  method requires only the objective
function, i.e.,  the logarithm of the joint likelihood, while the gradients
are computed numerically and the Hessian matrix of the second
order derivatives is updated in each iteration. The standard errors (SE) of the ML estimates can be also obtained via the gradients and the Hessian computed numerically during the maximization process.

\section{\label{blanketSec}Is the bivariate $K$-FNM a ``blanket" copula?}
In this section we will show that the $K$-FNM  copula is quite close to any parametric family of copulas. We will use the  Kullback-Leibler methodology \citep[pages 234-241]{joe2014} to compare the  new parametric copula family with existing parametric families of copulas. Before that, the  first subsection provides choices of parametric bivariate copulas.

\subsection{\label{bivcop}Existing parametric families of copulas}
We will consider copula families that have different tail dependence \citep{joe93} or tail order \citep{Hua-joe-11}.
A bivariate copula $C$ is {\it reflection symmetric}
if its density 
 satisfies $c(u_1,u_2)=c(1-u_1,1-u_2)$ for all $0\leq u_1,u_2\leq 1$.
Otherwise, it is reflection asymmetric often with more probability in the
joint upper tail or joint lower tail. {\it Upper tail dependence} means
that $c(1-u,1-u)=O(u^{-1})$ as $u\to 0$ and {\it lower tail dependence}
means that $c(u,u)=O(u^{-1})$ as $u\to 0$.
If $(U_1,U_2)\sim C$ for a bivariate copula $C$, then $(1-U_1,1-U_2)\sim
\widehat C$,  
where $\widehat C(u_1,u_2)=u_1+u_2-1+C(1-u_1,1-u_2)$   
is the survival or reflected 
copula of $C$; this ``reflection"
of each uniform $U(0,1)$ random variable about $1/2$ changes the direction
of tail asymmetry. Under some regularity conditions (e.g., existing finite density in the interior of the unit square, ultimately monotone in the tail), if there exists  $\kappa_L(C)>0$ and some $L(u)$ that is slowly varying at $0^+$ (i.e., $
 \frac{L(ut)}{L(u)} \sim 1,$ as $u\to 0^+$ for all  $t>0$), then  $\kappa_L(C)$  is the   \textit{lower tail order} of $C$. The \textit{upper tail order} $\kappa_U(C)$ can be defined by the reflection of $(U_1,U_2)$, i.e., $\Cbar(1-u,1-u) \sim u^{\kappa_U(C)} L^*(u)$ as $u\to0^+$, where $\Cbar$ is the survival function of the  copula and $L^*(u)$ is a slowly varying function. 
With $\kappa=\kappa_L$ or $\kappa_U$,  
 a bivariate copula has \textit{intermediate tail dependence} if $\kappa \in (1,2)$, \textit{tail dependence} if $\kappa=1$, and  \textit{tail quadrant independence} if $\kappa=2$ with $L(u)$ being asymptomatically a constant.

After briefly providing definitions of  tail dependence and  tail order we provide below a list of bivariate parametric copulas with varying tail behaviour: 
\begin{itemize}
\itemsep=10pt

\item Reflection symmetric copulas with intermediate tail dependence such as the BVN copula in (\ref{BVNcopula-cdf}) with $\kappa_L=\kappa_U=2/(1+ \th)$, where $\theta$ is the copula (correlation) parameter. 
\item Reflection symmetric copulas with tail quadrant independence ($\kappa_L=\kappa_U= 2$), such as the Frank copula.

\item Reflection asymmetric copulas with upper tail dependence only such as 
  the Gumbel copula with $\lambda_L=0$ ($\kappa_L=2^{1/\theta}$) and $\lambda_U=2^{1/\th}$ ($\kappa_U=1$), where $\theta$ is the copula parameter.  
  
  \item Reflection asymmetric copulas with lower tail dependence only such as 
  the Clayton copula with $\lambda_L=2^{-1/\th}$ ($\kappa_L=1$) and $\lambda_U=0$ ($\kappa_U=2$), where $\theta$ is the copula parameter.

\item Reflection symmetric copulas with tail dependence, such as the
$t_\nu$ copula with $\lambda=\lambda_L=\lambda_U=2 T_{\nu+1} \Bigl(- \sqrt{(\nu+1)(1 - \theta)/(1 + \theta)}\Bigr)$ ($\kappa_L=\kappa_U=1$), where $\theta$ is the correlation parameter of the bivariate $t$ distribution with $\nu$ degrees of freedom, and $T_{\nu}$ is the univariate $t$ cdf with $\nu$ degrees of freedom.  

\item Reflection asymmetric copulas with upper and lower tail dependence that can range independently from 0 to 1,  such as the BB1 copula with $\lambda_L=2^{-1/(\th\de)}$ ($\kappa_L=1$) and $\lambda_U=2-2^{1/\de}$ ($\kappa_U=1$) or the BB7 copula  with $\lambda_L=2^{-1/\de}$ ($\kappa_L=1$) and $\lambda_U=2^{1/\th}$ ($\kappa_U=1$), where $\theta$ and $\de$  are the copula parameters.

\end{itemize}

The aforementioned bivariate copula families are sufficient for applications because tail dependence and tail order  are  properties to consider when choosing amongst different families of copulas, and the concepts of upper/lower tail dependence and upper/lower tail order  are one way to differentiate families. \cite{Nikoloulopoulos&karlis08CSDA} and \cite{joe2014} have shown that it is hard to choose a copula with similar tail dependence properties from real data because copulas with similar tail dependence properties provide similar fit.

\subsection{Kullback-Leibler distance and sample size}

For inferences based on likelihood, the Kullback-Leibler (KL) distance is relevant, especially as the parametric model used in the likelihood could be misspecified \citep{joe2014}. 
Typically, one considers several different models when analysing data, and from a theoretical point of view, the KL distance of pairs of competing models provides information on the sample size needed to discriminate them.

We will define the KL distance for two copula densities and the expected log-likelihood ratio. Because the KL  is a non-negative quantity that is not bounded above,  we use also the expected value of the square of the log-likelihood ratio in order to get a sample size value that is an indication of how different two copula densities are.
Consider two  copula densities (competing models) $c_1$ and $c_2$ with respect to Lebesgue or counting measure in $\mathbb{R}^2$.  The  KL distance between copulas with densities $c_1,c_2$
is defined  as
\begin{equation}\label{KLdistance}
  \mbox{KL}(c_1,c_2) = E_{c_1} \left[ \log \left( \frac{c_1(u_1,u_2)}{c_2(u_1,u_2)}\right) \right] =
  \int_{0}^{1}\int_{0}^{1} \log \left( \frac{c_1(u_1,u_2)}{c_2(u_1,u_2)}\right) c_1(u_1,u_2)
  du_1du_2.
\end{equation}
 The  KL distance can be interpreted as the average difference of the contribution to the
  log-likelihood of one observation.

We use the log-likelihood ratio to get a sample size $n_{c_1,c_2}$ which gives an indication of the sample size needed to distinguish $c_1$ and $c_2$ with probability at least 0.95. If $c_1,c_2$ are similar, then $c_1,c_2$  will be larger, and if $c_1,c_2$ are far apart, then $n_{c_1,c_2}$  will be smaller. The calculation is based on an approximation from the Central Limit Theorem and assumes that the square of the log-likelihood ratio has finite variance when computed with $c_1$  being the true density \citep{joe2014}.
If the variance of the log-density ratio is
$$
\sigma_{c_1}^2=
 \int_{0}^{1}\int_{0}^{1} \left[\log \left( \frac{c_1(u_1,u_2)}{c_2(u_1,u_2)}\right)\right]^2 c_1(u_1,u_2)
  du_1du_2- \Bigl[\mbox{KL}(c_1,c_2)\Bigr]^2,
$$
then the KL sample size   is 
$$n_{c_1,c_2}=
\Phi^{-1}(0.95)\left[\frac{\sigma_{c_1}}{\mbox{KL}(c_1,c_2)}\right]^2. 
$$
This is larger when $\mbox{KL}(c_1,c_2)$ is small or  the variance   $\sigma_{c_1}^2$ is large.

\subsection{Minimizing the KL distance}

For a theoretical likelihood comparison between existing bivariate parametric families of copulas 
and the bivariate  $K$-FNM copula  we minimize  the KL
distance in (\ref{KLdistance}) where the true $c_1$ is the copula density of each of  parametric bivariate copulas  in Subsection \ref{bivcop} and $c_2$ is the copula density of the $K$-FNM copula in (\ref{FNMdensity}),  and hence (a) obtain the  parameters of the $K$-FNM copula that is quite close to the true copula in KL distance, (b) the KL sample size for these parameters.
The minimized  KL distances and resultant sample sizes will show the similarity or dissimilarity of the $K$-FNM copula with the existing parametric families of copulas.

Numerical evaluation of $\mbox{KL}(c_1,c_2)$ or  the variance   $\sigma_{c_1}^2$   can be approached using dependent Gauss-Legendre quadrature points \citep{Nikoloulopoulos2015b} with the following steps:

\begin{enumerate}
\itemsep=0pt
\item Calculate Gauss-Legendre  quadrature points $\{u_q: q=1,\ldots,n_q\}$ 
and weights $\{w_q: q=1,\ldots,n_q\}$ in terms of standard uniform; see e.g., \cite{Stroud&Secrest1966}.
\item Convert from independent uniform random variables $\{u_{q_1}: q_1=1,\ldots,n_q\}$ and $\{u_{q_2}: q_2=1,\ldots,n_q\}$ to dependent uniform random variables $\{u_{q_1}: q_1=1,\ldots,n_q\}$ and $\{C_1^{-1}(u_{q_2}|u_{q_1};\th): q_1=q_2=1,\ldots,n_q\}$ that have copula  $C_1$.
The inverse of the conditional distribution $C_1(u_2|u_1)=\partial C_1(u_1,u_2)/\partial u_1$ corresponding to the copula $C_1$ is used  to achieve this.
\item Numerically evaluate 
$$\int_{0}^{1}\int_{0}^{1} \left[\log \left( \frac{c_1(u,v)}{c_2(u,v)}\right)\right]^p c_1(u,v)
  dudv \quad \mbox{for} \quad p=1,2$$

\noindent in a double sum:
$$\mathlarger{\mathlarger{\mathlarger{‎‎\sum}}}_{q_1=1}^{n_q}\mathlarger{\mathlarger{\mathlarger{‎‎\sum}}}_{q_2=1}^{n_q}w_{q_1}w_{q_2}
\left[\log \left( \frac{c_1\Bigr(u_{q_1},C_1^{-1}(u_{q_2}|u_{q_1})\Bigl)}{c_2\Bigr(u_{q_1},C_1^{-1}(u_{q_2}|u_{q_1})\Bigl)}\right)\right]^p.$$
\end{enumerate}
With Gauss-Legendre quadrature, the same nodes and weights
are used for different functions;
this helps in yielding smooth numerical derivatives for numerical optimization via quasi-Newton \citep{nash90}.
Our comparisons show that $n_q=15$ is adequate with good precision to at least at four decimal places; hence it also provides the advantage of fast implementation.

\setlength{\tabcolsep}{10pt}
\begin{table}[!t]
  \centering
  \caption{\label{1-par-comparison}Minimized  KL distances and the corresponding 2-FNM copula parameters and KL sample sizes for comparing 1-parameter  copula families, with symmetric or asymmetric  dependence  as the Kendall's $\tau$ varies from 0.1 to 0.9, versus the bivariate 2-FNM copula.}
    \begin{tabular}{cccccccccc}
    \toprule
Copula&    $\tau$ & $\lambda_L$ & $\lambda_U$ & $\mbox{KL}(c_1,c_2)$    & $\pi$ & $\th$ & $\rho_1$ & $\rho_2$ & $n_{c_1,c_2}$ \\

    \midrule
    BVN   & 0.1   & 0     & 0     & 0.000 & 0.500 & 0.090 & 0.134 & 0.134 & 5482651 \\
          & 0.2   & 0     & 0     & 0.000 & 0.500 & 0.179 & 0.267 & 0.267 & 248291 \\
          & 0.3   & 0     & 0     & 0.000 & 0.500 & 0.269 & 0.398 & 0.398 & 47223 \\
          & 0.4   & 0     & 0     & 0.000 & 0.500 & 0.359 & 0.526 & 0.526 & 16696 \\
          & 0.5   & 0     & 0     & 0.000 & 0.500 & 0.451 & 0.647 & 0.647 & 8508 \\
          & 0.6   & 0     & 0     & 0.001 & 0.500 & 0.544 & 0.758 & 0.758 & 5287 \\
          & 0.7   & 0     & 0     & 0.001 & 0.500 & 0.642 & 0.855 & 0.855 & 3475 \\
          & 0.8   & 0     & 0     & 0.002 & 0.500 & 0.746 & 0.931 & 0.931 & 2301 \\
          & 0.9   & 0     & 0     & 0.003 & 0.493 & 0.864 & 0.982 & 0.982 & 1500 \\\hline
    Frank & 0.1   & 0     & 0     & 0.000 & 0.500 & 0.149 & 0.060 & 0.060 & 14417 \\
          & 0.2   & 0     & 0     & 0.002 & 0.500 & 0.306 & 0.122 & 0.122 & 3729 \\
          & 0.3   & 0     & 0     & 0.003 & 0.500 & 0.483 & 0.188 & 0.188 & 1814 \\
          & 0.4   & 0     & 0     & 0.005 & 0.500 & 0.680 & 0.274 & 0.274 & 1175 \\
          & 0.5   & 0     & 0     & 0.007 & 0.500 & 0.884 & 0.396 & 0.396 & 837 \\
          & 0.6   & 0     & 0     & 0.011 & 0.500 & 1.096 & 0.549 & 0.549 & 537 \\
          & 0.7   & 0     & 0     & 0.024 & 0.500 & 1.400 & 0.715 & 0.715 & 260 \\
          & 0.8   & 0     & 0     & 0.067 & 0.500 & 1.161 & 0.860 & 0.860 & 111 \\
          & 0.9   & 0     & 0     & 0.178 & 0.505 & 1.048 & 0.955 & 0.954 & 65 \\\hline
    Clayton & 0.1   & 0.04  & 0     & 0.001 & 0.964 & 0.726 & 0.094 & -0.116 & 5262 \\
          & 0.2   & 0.25  & 0     & 0.003 & 0.917 & 0.794 & 0.171 & 0.285 & 1836 \\
          & 0.3   & 0.45  & 0     & 0.005 & 0.869 & 0.863 & 0.248 & 0.591 & 1032 \\
          & 0.4   & 0.59  & 0     & 0.008 & 0.812 & 0.915 & 0.326 & 0.762 & 669 \\
          & 0.5   & 0.71  & 0     & 0.012 & 0.748 & 0.957 & 0.411 & 0.858 & 420 \\
          & 0.6   & 0.79  & 0     & 0.020 & 0.687 & 0.991 & 0.519 & 0.921 & 247 \\
          & 0.7   & 0.86  & 0     & 0.034 & 0.618 & 1.008 & 0.634 & 0.960 & 140 \\
          & 0.8   & 0.92  & 0     & 0.061 & 0.538 & 1.010 & 0.755 & 0.983 & 79 \\
          & 0.9   & 0.96 & 0     & 0.400 & 0.779 & 0.978 & 0.972 & 0.990 & 51 \\\hline
    Gumbel & 0.1   & 0     & 0.13  & 0.000 & 0.012 & 1.075 & 0.638 & 0.129 & 104054 \\
          & 0.2   & 0     & 0.26  & 0.002 & 0.026 & 0.992 & 0.642 & 0.254 & 3962 \\
          & 0.3   & 0     & 0.38  & 0.004 & 0.049 & 0.946 & 0.673 & 0.368 & 1891 \\
          & 0.4   & 0     & 0.48  & 0.005 & 0.079 & 0.929 & 0.755 & 0.481 & 1405 \\
          & 0.5   & 0     & 0.59  & 0.006 & 0.109 & 0.938 & 0.833 & 0.596 & 1153 \\
          & 0.6   & 0     & 0.68  & 0.006 & 0.143 & 0.953 & 0.894 & 0.708 & 1004 \\
          & 0.7   & 0     & 0.77  & 0.007 & 0.181 & 0.970 & 0.940 & 0.814 & 877 \\
          & 0.8   & 0     & 0.85  & 0.008 & 0.221 & 0.986 & 0.973 & 0.905 & 758 \\
          & 0.9   & 0     & 0.93  & 0.019 & 0.466 & 0.985 & 0.990 & 0.958 & 258 \\
    \bottomrule
    \end{tabular}%
\end{table}

Table \ref{1-par-comparison} shows the minimized  KL distances and the corresponding 2-FNM copula parameters and KL sample sizes for comparing 1-parameter  copula families, with symmetric or asymmetric  dependence  as the Kendall's $\tau$ varies from 0.1 to 0.9, versus the bivariate 2-FNM copula. 
Table \ref{bb1} shows the minimized  KL distances and the corresponding 2-FNM  or 3-FNM copula parameters and KL sample sizes for comparing the BB1 copula, with reflection  asymmetric  tail dependence ($\lambda_L\neq \lambda_U$) as the lower and upper  tail dependence  varies from 0.1 to 0.9 and from 0.9 to 0.1,  respectively, versus the bivariate 2- or 3-FNM copula.
 Table \ref{t-copula-comparisons} shows the minimized  KL distances and the corresponding 2-FNM or 3-FNM copula parameters and KL sample sizes  for comparing the $t_\nu$ copula for a small $\nu$, with reflection symmetric tail dependence ($\lambda_L=\lambda_U$) as the Kendall's $\tau$ varies from 0.1 to 0.9, versus the bivariate 2- or 3-FNM copula.

\setlength{\tabcolsep}{7pt}
\begin{table}[!h]
  \centering
  \caption{\label{bb1}Minimized  KL distances and the corresponding 2-FNM  or 3-FNM copula parameters and KL sample sizes for comparing the BB1 copula, with reflection  asymmetric  tail dependence ($\lambda_L\neq \lambda_U$) as the lower and upper  tail dependence  varies from 0.1 to 0.9 and from 0.9 to 0.1,  respectively, versus the bivariate 2- or 3-FNM copula.}
    \begin{tabular}{ccccccccccccc}
    \toprule
     $\lambda_L$ & $\lambda_U$ &$\tau$ & $K$ & $\mbox{KL}(c_1,c_2)$   & $\pi_1$ & $\pi_2$ & $\th_1$ & $\th_2$ & $\rho_1$ & $\rho_2$ & $\rho_3$ & $n_{c_1,c_2}$ \\
    \midrule
    0.1   & 0.9   & 0.87  & 2     & 0.009 & 0.227 &       & 0.994 &       & 0.988 & 0.956 &       & 737 \\
    0.2   & 0.8   & 0.75  & 3     & 0.006 & 0.087 & 0.426 & 1.599 & -0.799 & 0.918 & 0.951 & 0.850 & 1271 \\
    0.3   & 0.7   & 0.66  & 2     & 0.006 & 0.081 &       & 0.978 &       & 0.932 & 0.821 &       & 1140 \\
    0.4   & 0.6   & 0.59  & 3     & 0.001 & 0.026 & 0.545 & 1.729 & -0.858 & 0.576 & 0.695 & 0.894 & 4437 \\
    0.5   & 0.5   & 0.55  & 3     & 0.005 & 0.005 & 0.537 & 3.565 & -1.765 & 0.670 & 0.654 & 0.881 & 909 \\
    0.6   & 0.4   & 0.54  & 2     & 0.007 & 0.923 &       & 0.986 &       & 0.687 & 0.853 &       & 963 \\
    0.7   & 0.3   & 0.56  & 2     & 0.006 & 0.839 &       & 0.973 &       & 0.642 & 0.889 &       & 979 \\
    0.8   & 0.2   & 0.63  & 2     & 0.013 & 0.728 &       & 0.994 &       & 0.643 & 0.934 &       & 414 \\
    0.9   & 0.1   & 0.77  & 2     & 0.042 & 0.580 &       & 1.009 &       & 0.743 & 0.977 &       & 120 \\
    \bottomrule
    \end{tabular}%
\end{table}

The conclusion from the values in the tables are:
\begin{itemize}
\itemsep=10pt
\item The $K$-FNM copula  is close to any parametric bivariate family of copulas and a large sample size is required to distinguish when the Kendall's $\tau$ values range from 0.1 (weak dependence) to 0.5 (moderate dependence). 
\item To approximate copulas with refection symmetric or asymmetric  tail dependence, they are required up to $K=3$ mixture components, while for any  1-parameter family $K=2$ mixture components are sufficient. 

\item Since the $K$-FNM copula and each of the parametric families of copulas have the same strength of dependence as given by Kendall's $\tau$,   the magnitude of the KL distance is related to the closeness of the strength of dependence in the tails. This is because copula densities can asymptote to infinity in a joint tail at different rates (tail order less than dimension d) or converge to a constant in the joint tail (if tail order is the dimension d or larger), see e.g,  \cite{joe2014}.

\item Copula families with stronger dependence have larger KL distance  with the $K$-FNM copula than those with weaker dependence when strength of dependence in the tails are different based on the tail orders. 
\end{itemize}

\begin{landscape}
\begin{figure}[!h]
\begin{center}
\begin{tabular}{cc|cc}
\hline
\multicolumn{2}{c|}{$\tau=0.5 \quad \pi_1=0.5 \quad \th_1=0.45 \quad \rho_1=0.65\quad\rho_2=0.65$}&\multicolumn{2}{c}{ $\tau=0.5 \quad  \pi_1= 0.5 \quad \th_1= 0.88 \quad \rho_1=0.40\quad \rho_2=0.40$}\\\hline
\includegraphics[width=0.3\textwidth]{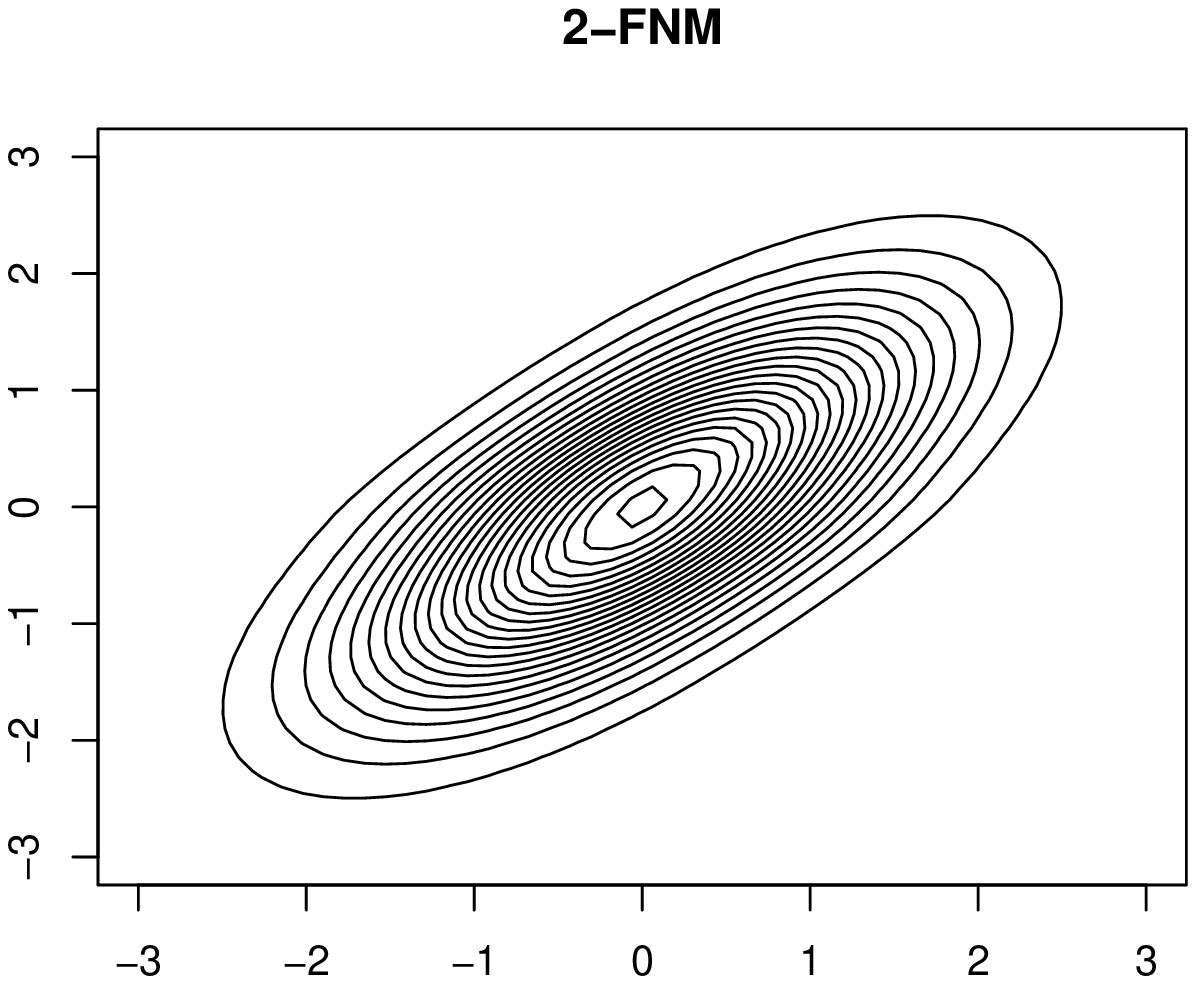} &
\includegraphics[width=0.3\textwidth]{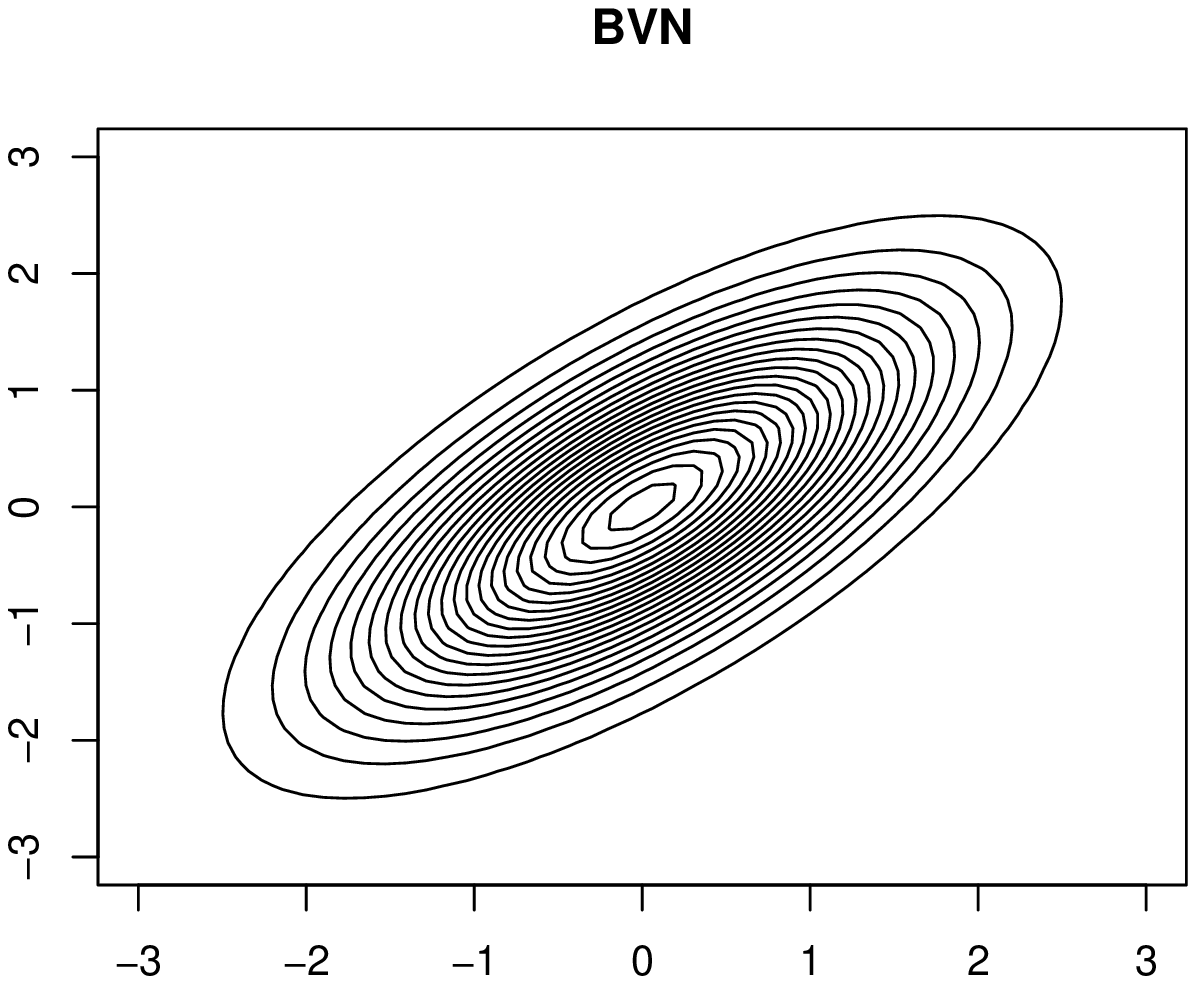}&
\includegraphics[width=0.3\textwidth]{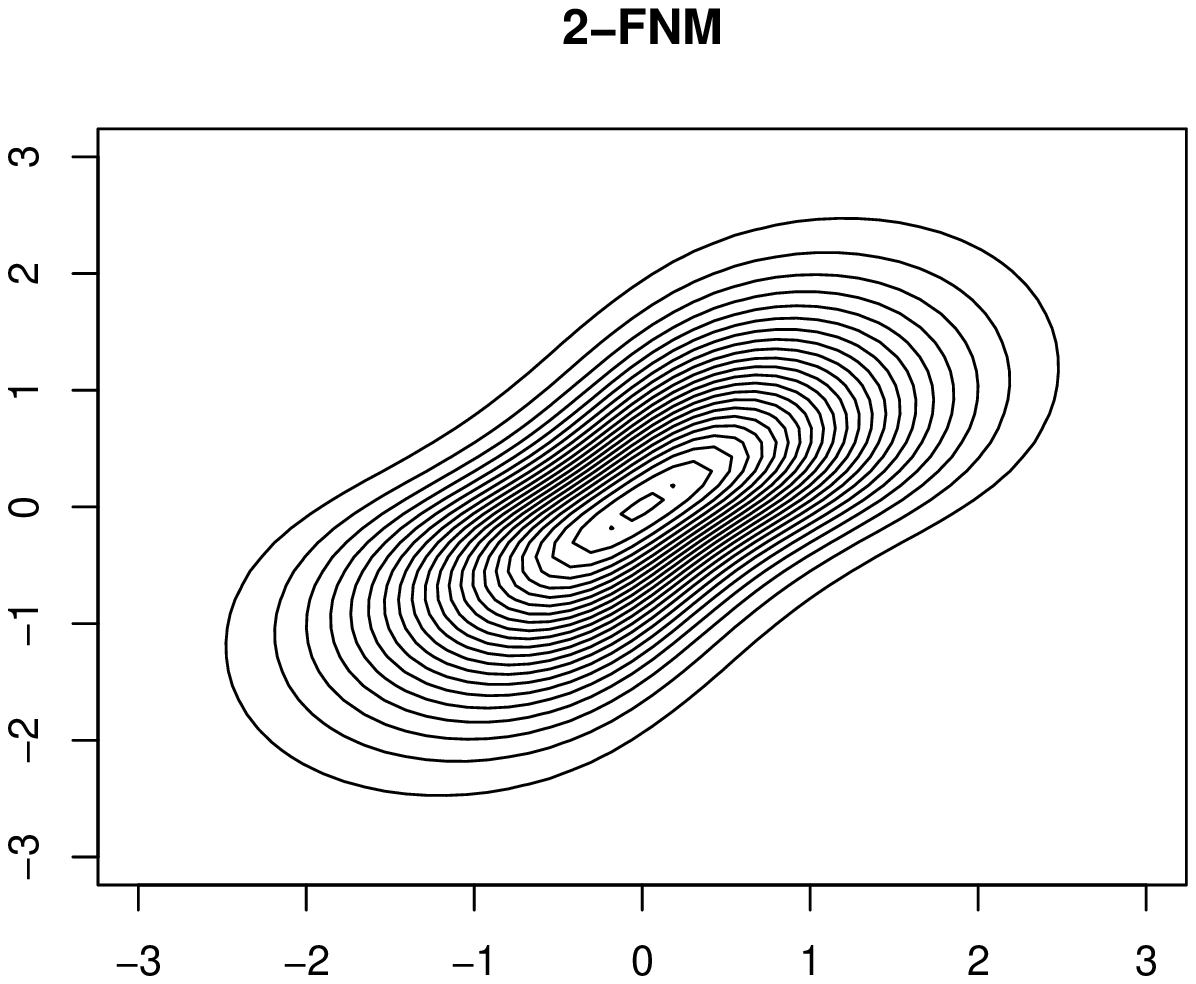} &
\includegraphics[width=0.3\textwidth]{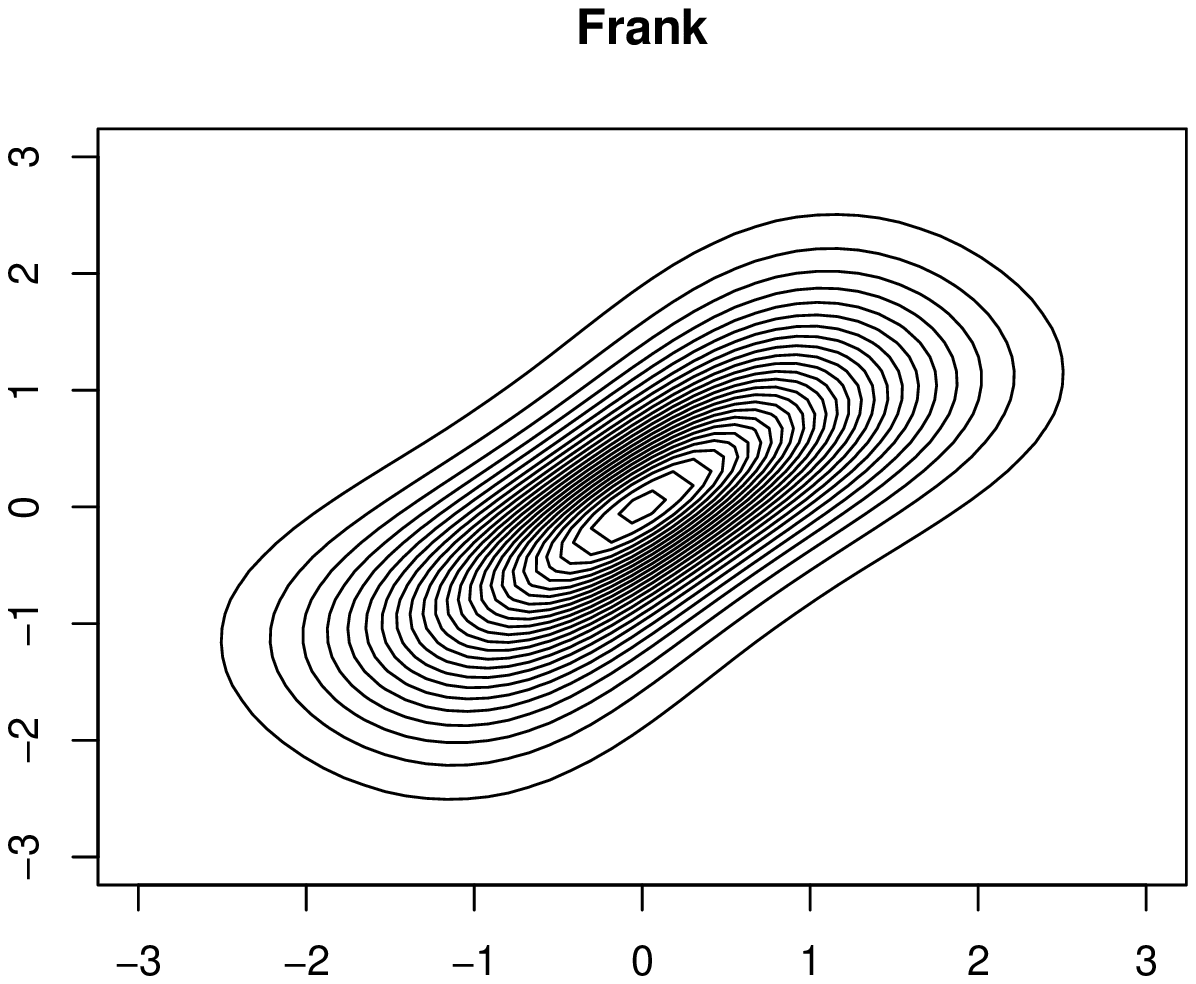}\\
\hline
\multicolumn{2}{c|}{$\tau=0.5 \quad \pi_1=0.75 \quad \th_1=0.96 \quad \rho_1=0.41 \quad \rho_2= 0.86$}&\multicolumn{2}{c}{ $\tau=0.5 \quad  \pi_1=0.11 \quad \th_1= 0.94 \quad \rho_1=0.83 \quad \rho_2 =0.60 $}\\\hline
\includegraphics[width=0.3\textwidth]{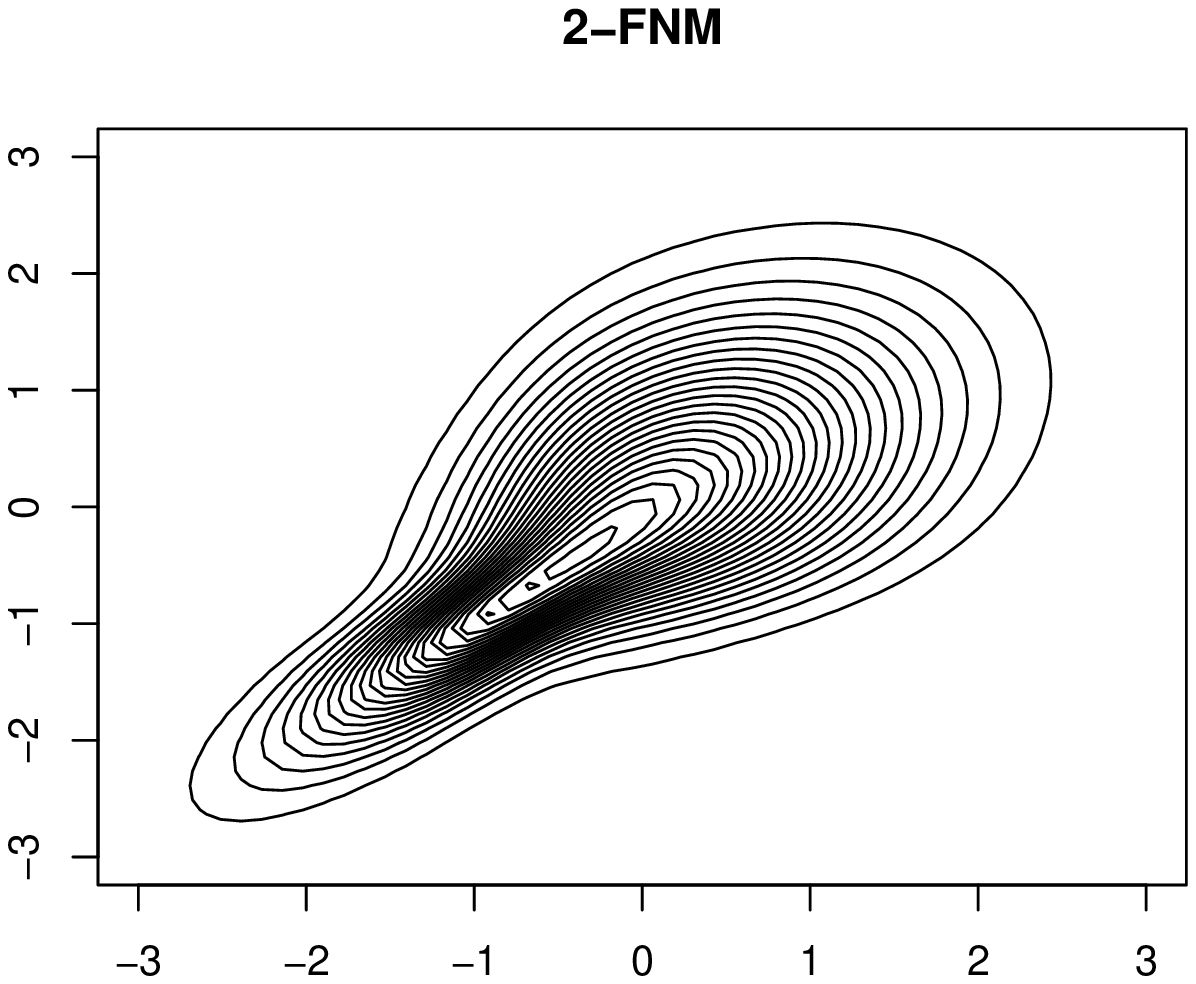} &
\includegraphics[width=0.3\textwidth]{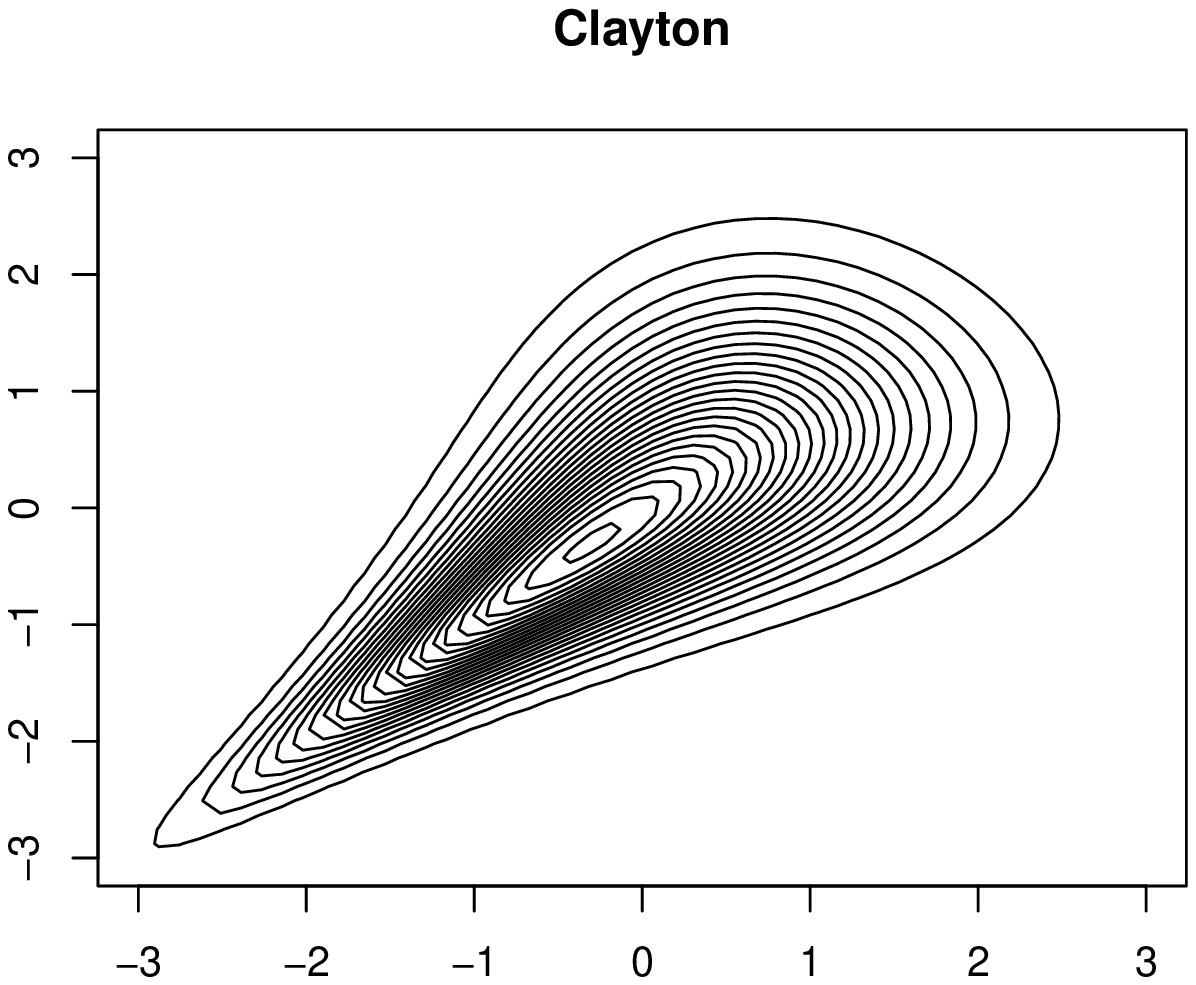}
&\includegraphics[width=0.3\textwidth]{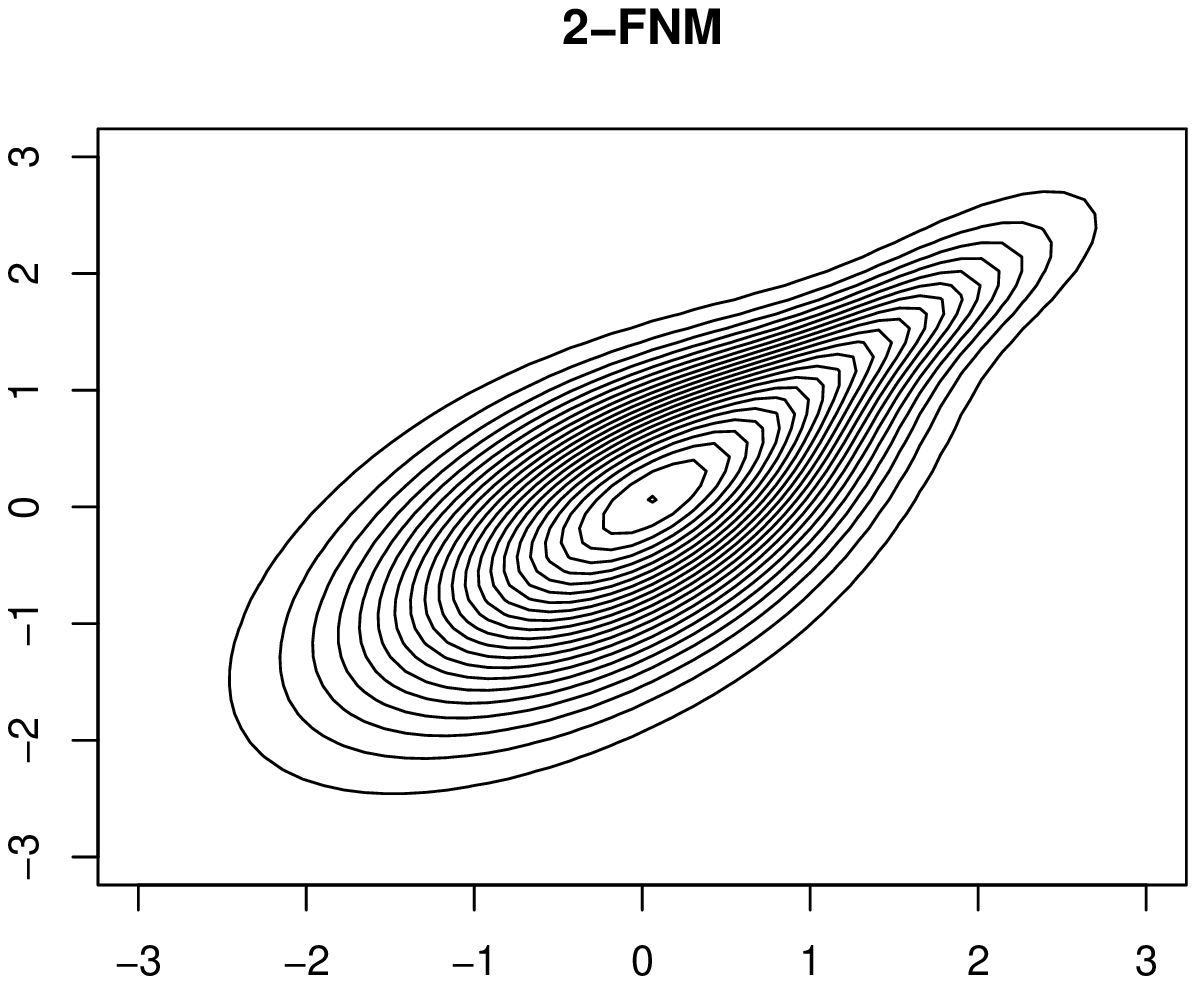} &
\includegraphics[width=0.3\textwidth]{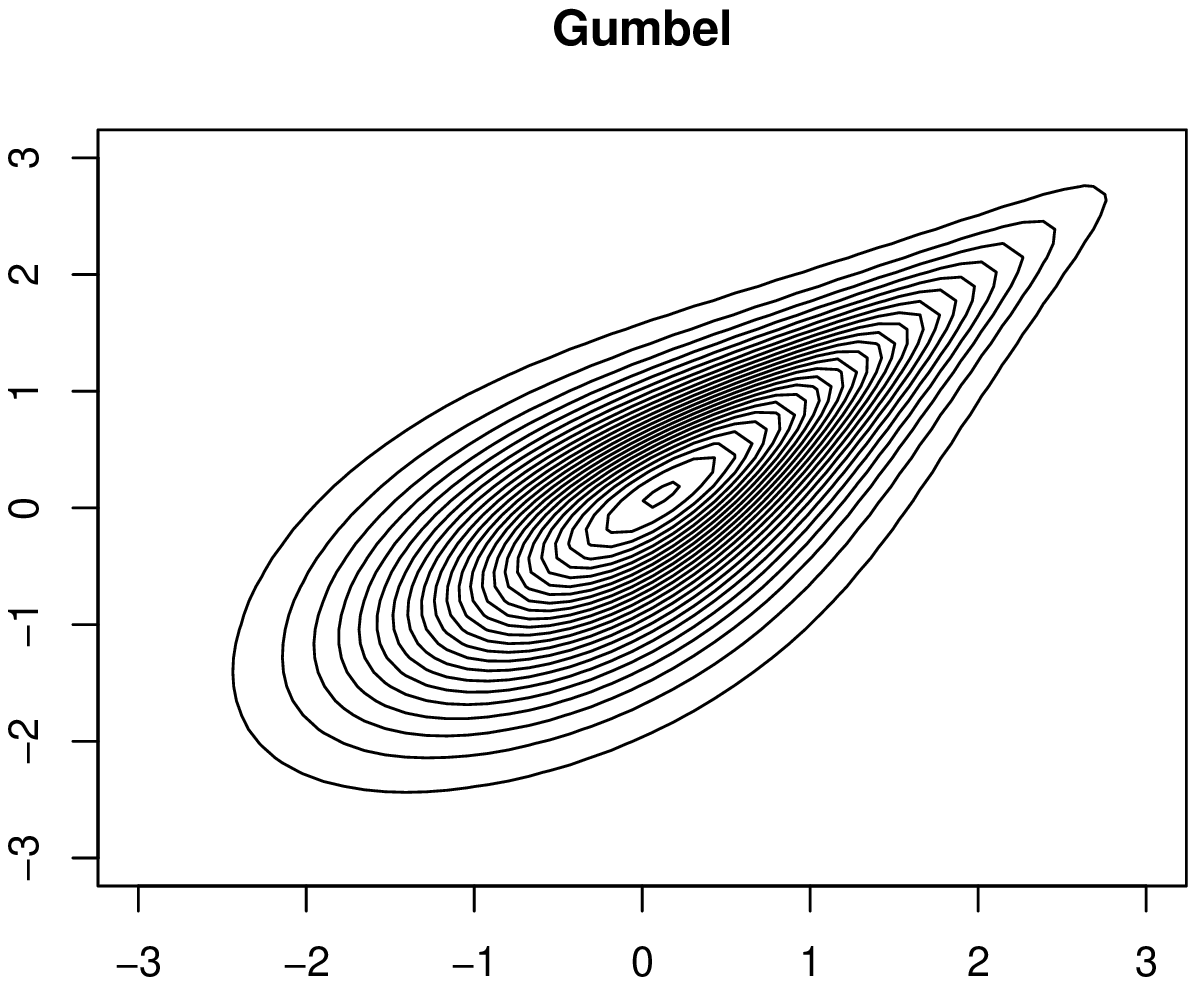}\\
\hline
\multicolumn{2}{c|}{$\tau=0.5 \quad  \pi_1=0.01 \quad \pi_2=0.17 \quad \th_1= 1.65 $}&\multicolumn{2}{c}{ $\lambda_L=0.4  \quad  \lambda_U=0.6   \quad \tau= 0.59  \quad \pi_1=0.03 \quad \pi_2=0.55
$}\\
\multicolumn{2}{c|}{$\th_2= -0.81 \quad \rho_1= 0.71  \quad \rho_2=-0.35  \quad \rho_3= 0.85$}&\multicolumn{2}{c}{ $\th_1=1.73 \quad \th_3=-0.86  \quad \rho_1=0.58 \quad \rho_2=0.69\quad  \rho_3=0.89
$}\\\hline
\includegraphics[width=0.3\textwidth]{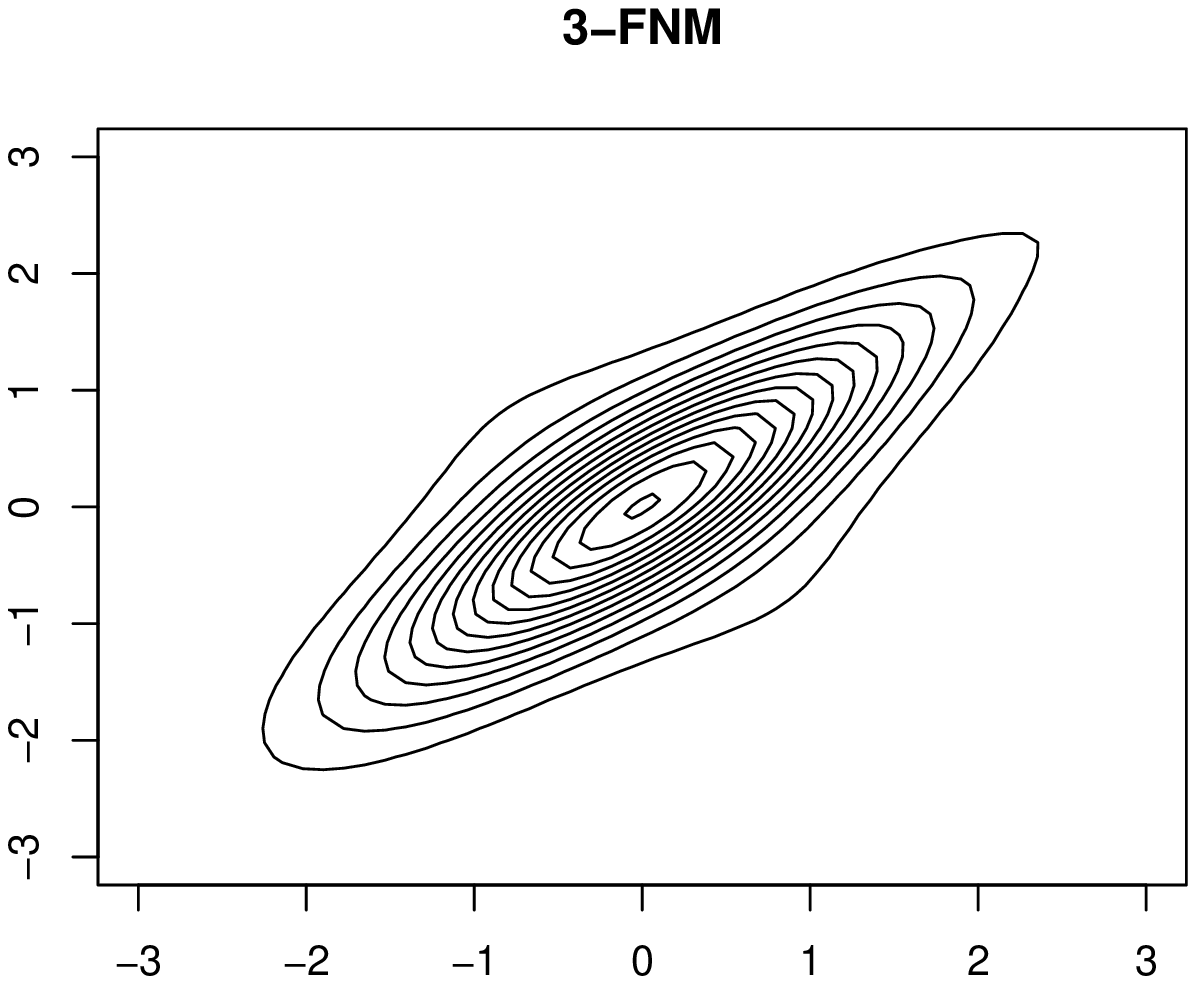} &
\includegraphics[width=0.3\textwidth]{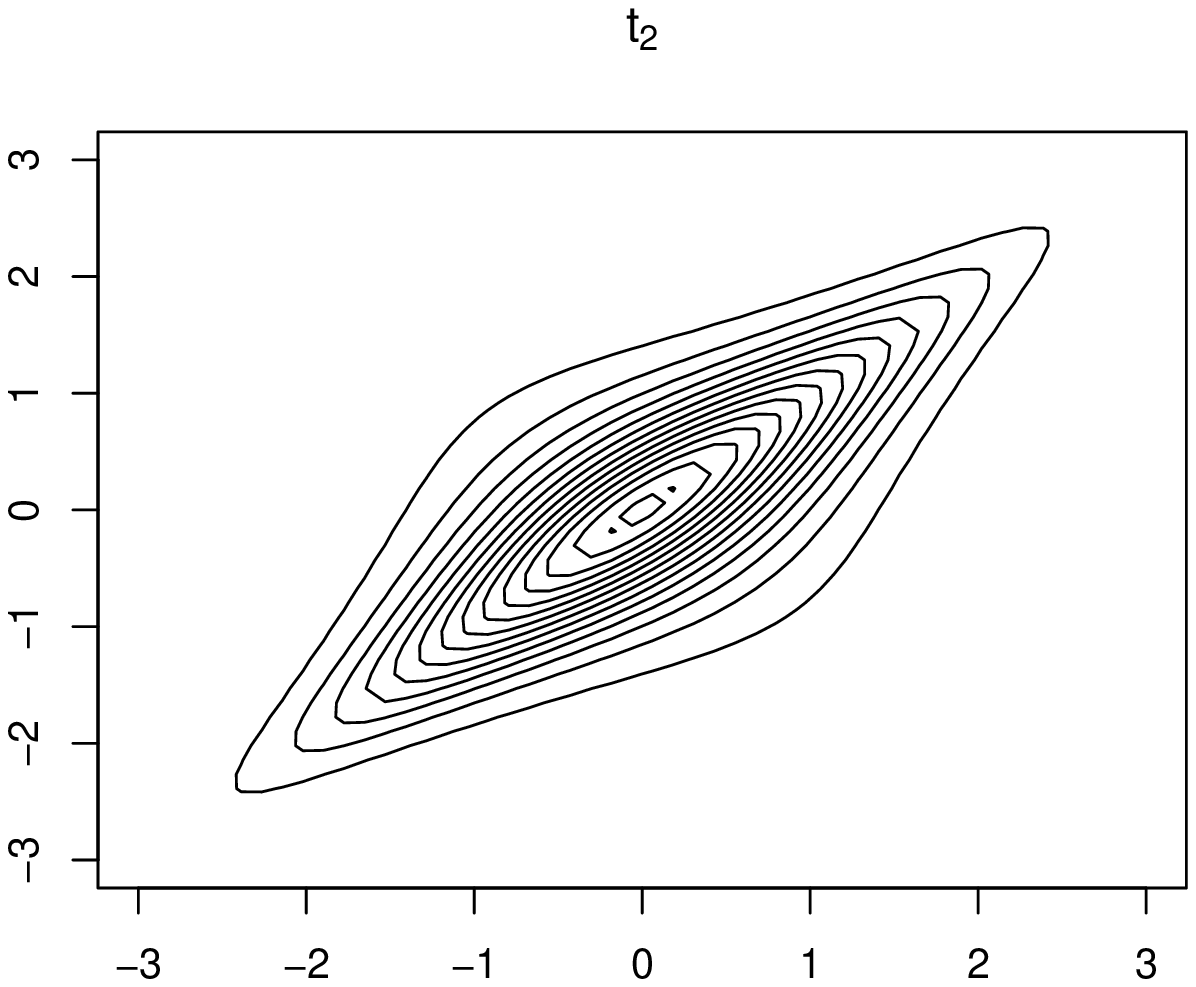}&
\includegraphics[width=0.3\textwidth]{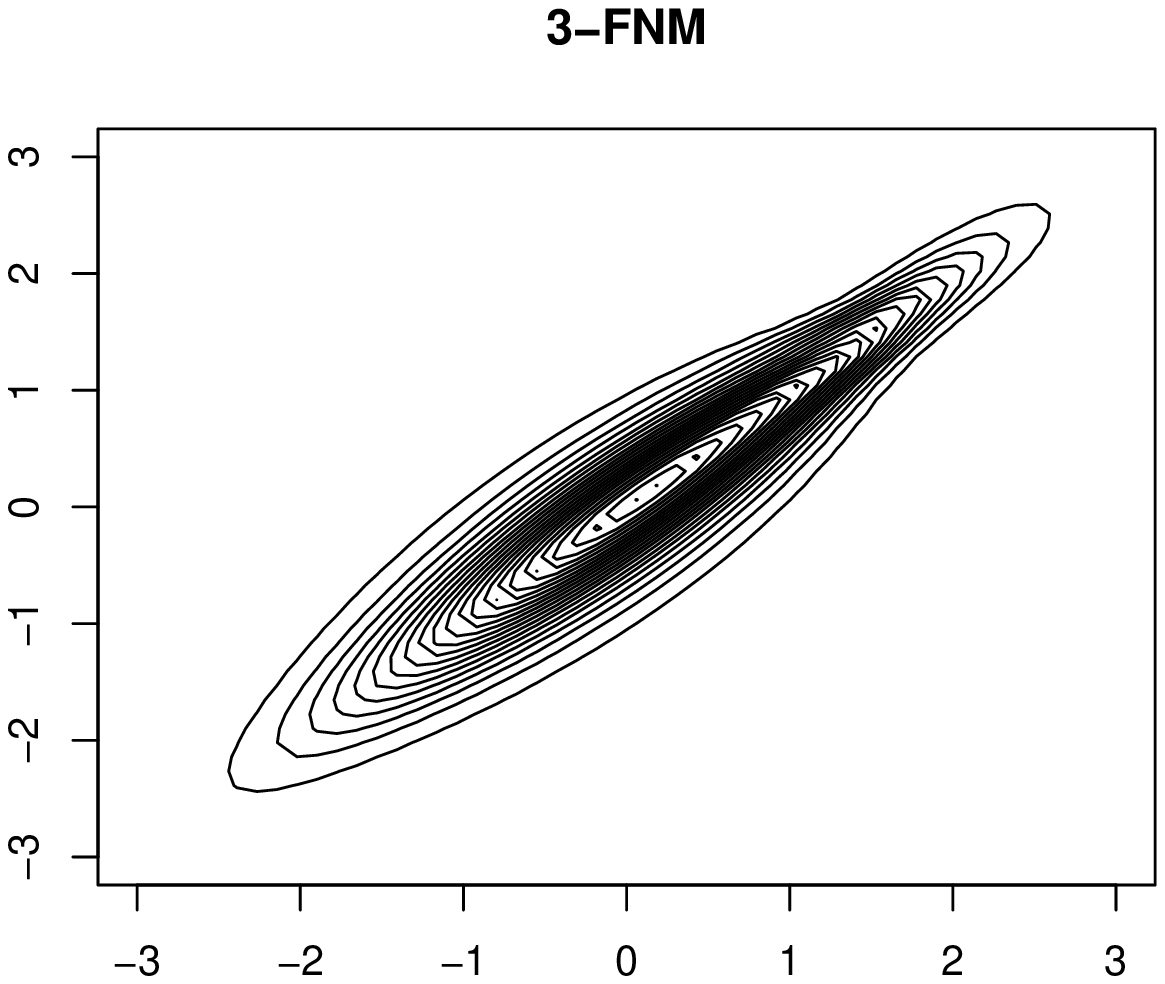} &
\includegraphics[width=0.3\textwidth]{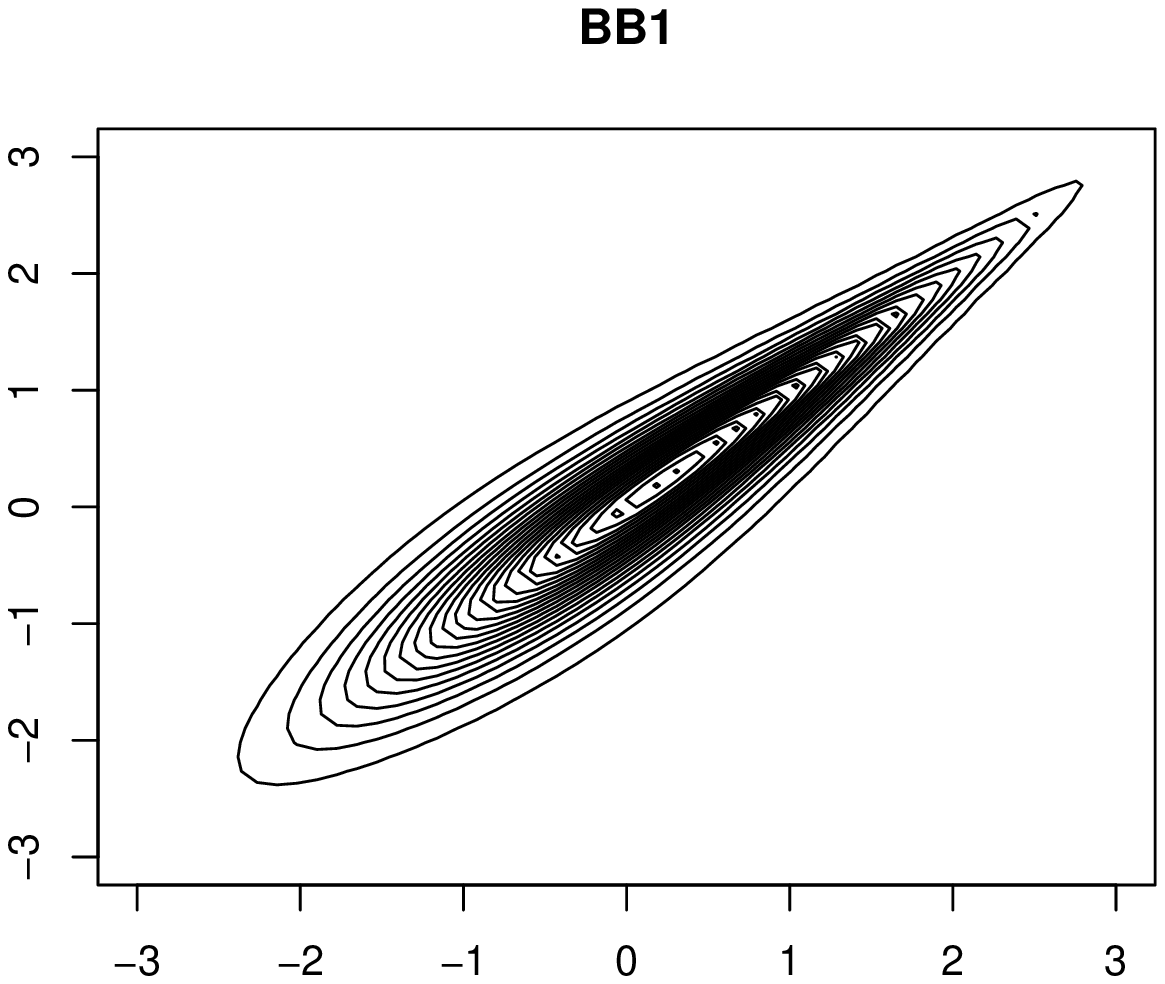}\\
\hline
\end{tabular}

\caption{\label{summary-figure}Contour plots of the 2- or 3-FNM copula with the parameters in Tables   \ref{1-par-comparison}--\ref{t-copula-comparisons}, i.e., the ones that the FNM copulas are  close in terms of KL distance to the true copulas, and normal margins, along with the contour plots of the true copulas  with normal margins.  }
\end{center}
\end{figure}
\end{landscape}

Figure \ref{summary-figure} summarizes these results by depicting the contour plots of the 2- or 3-FNM copula with the parameters in Tables   \ref{1-par-comparison}--\ref{t-copula-comparisons}, i.e., the ones that the FNM copulas are  close in terms of KL distance to the true copulas, and normal margins, along with the contour plots of the true copulas with normal margins.    We summarize the case of $\tau=0.5$ ($\lambda_L=0.4,\lambda_U=0.6$ for BB1).

\setlength{\tabcolsep}{5pt}
\begin{table}[!h]
  \centering
  \caption{\label{t-copula-comparisons}Minimized  KL distances and the corresponding 2-FNM or 3-FNM copula parameters and KL sample sizes  for comparing the $t_\nu$ copula for a small $\nu$, with reflection symmetric tail dependence ($\lambda_L=\lambda_U$) as the Kendall's $\tau$ varies from 0.1 to 0.9, versus the bivariate 2- or 3-FNM copula.}
    \begin{tabular}{cccccccccccccc}
    \toprule
    $\nu$ & $\tau$ & $\lambda_L$ & $\lambda_U$ & $K$ & $\mbox{KL}(c_1,c_2)$   & $\pi_1$ & $\pi_2$ & $\th_1$ & $\th_2$ & $\rho_1$ & $\rho_2$ & $\rho_3$ & $n_{c_1,c_2}$ \\
    \midrule
    2     & 0.1   & 0.24  & 0.24  & 3     & 0.004 & 0.004 & 0.436 & 2.003 & -0.986 & 0.756 & -0.582 & 0.689 & 1602 \\
          & 0.2   & 0.30  & 0.30  & 3     & 0.004 & 0.005 & 0.365 & 1.988 & -0.978 & 0.738 & -0.531 & 0.731 & 1535 \\
          & 0.3   & 0.37  & 0.37  & 3     & 0.004 & 0.005 & 0.291 & 2.037 & -1.004 & 0.737 & -0.488 & 0.769 & 1338 \\
          & 0.4   & 0.44  & 0.44  & 3     & 0.006 & 0.006 & 0.221 & 1.978 & -0.972 & 0.732 & -0.443 & 0.808 & 1002 \\
          & 0.5   & 0.52  & 0.52  & 3     & 0.008 & 0.010 & 0.170 & 1.653 & -0.807 & 0.705 & -0.350 & 0.852 & 640 \\
          & 0.6   & 0.61  & 0.61  & 3     & 0.012 & 0.010 & 0.166 & 1.906 & -0.938 & 0.735 & 0.057 & 0.906 & 375 \\
          & 0.7   & 0.71  & 0.71  & 3     & 0.014 & 0.011 & 0.156 & 2.015 & -0.997 & 0.817 & 0.435 & 0.945 & 498 \\
          & 0.8   & 0.80  & 0.80  & 3     & 0.026 & 0.000 & 0.108 & 1.666 & -0.819 & 0.599 & 0.441 & 0.976 & 181 \\
          & 0.9   & 0.90  & 0.90  & 2     & 0.085 & 0.007 &       & 0.694 &       & 0.990 & 0.989 &       & 109 \\ \midrule
    3     & 0.1   & 0.16  & 0.16  & 3     & 0.001 & 0.003 & 0.411 & 2.103 & -1.037 & 0.688 & -0.498 & 0.589 & 10194 \\
          & 0.2   & 0.22  & 0.22  & 3     & 0.001 & 0.004 & 0.331 & 2.118 & -1.044 & 0.656 & -0.441 & 0.643 & 6923 \\
          & 0.3   & 0.29  & 0.29  & 3     & 0.002 & 0.004 & 0.264 & 2.137 & -1.053 & 0.638 & -0.363 & 0.703 & 3576 \\
          & 0.4   & 0.37  & 0.37  & 3     & 0.003 & 0.005 & 0.213 & 2.318 & -1.143 & 0.626 & -0.234 & 0.766 & 1865 \\
          & 0.5   & 0.45  & 0.45  & 3     & 0.005 & 0.005 & 0.185 & 6.257 & -3.115 & 0.691 & 0.004 & 0.830 & 1050 \\
          & 0.6   & 0.55  & 0.55  & 3     & 0.006 & 0.007 & 0.195 & 2.103 & -1.038 & 0.697 & 0.365 & 0.892 & 966 \\
          & 0.7   & 0.66  & 0.66  & 3     & 0.006 & 0.008 & 0.206 & 2.085 & -1.035 & 0.801 & 0.650 & 0.940 & 983 \\
          & 0.8   & 0.77  & 0.77  & 3     & 0.023 & 0.000 & 0.051 & 0.265 & -0.290 & 0.302 & 0.279 & 0.966 & 174 \\
          & 0.9   & 0.88  & 0.88  & 3     & 0.050 & 0.001 &  0.000& 0.761 & 0.621 & 0.154 & 0.171 & 0.987 & 123 \\ \midrule
    4     & 0.1 & 0.11  & 0.11  & 3     & 0.000 & 0.001 & 0.403 & 2.609 & -1.302 & -0.990 & -0.432 & 0.543 & 34841 \\
          & 0.2   & 0.17  & 0.17  & 3     &0.000  & 0.003 & 0.315 & 2.372 & -1.172 & 0.626 & -0.369 & 0.593 & 119419 \\
          & 0.3 & 0.23  & 0.23  & 3     & 0.001 & 0.003 & 0.247 & 2.683 & -1.326 & 0.545 & -0.276 & 0.662 & 15265 \\
          & 0.4   & 0.31  & 0.31  & 3     & 0.002 & 0.005 & 0.213 & 2.858 & -1.409 & 0.545 & -0.085 & 0.739 & 4058 \\
          & 0.5   & 0.40  & 0.40  & 3     & 0.003 & 0.005 & 0.222 & 2.609 & -1.288 & 0.611 & 0.229 & 0.820 & 2052 \\
          & 0.6   & 0.50  & 0.50  & 3     & 0.003 & 0.006 & 0.259 & 2.596 & -1.287 & 0.675 & 0.531 & 0.890 & 1935 \\
          & 0.7   & 0.61  & 0.61  & 3     & 0.002 & 0.007 & 0.266 & 1.981 & -0.985 & 0.775 & 0.738 & 0.938 & 3018 \\
          & 0.8   & 0.74  & 0.74  & 2     & 0.025 & 0.951 &       & 0.925 &       & 0.949 & 0.848 &       & 256 \\
          & 0.9   & 0.87  & 0.87  & 2     & 0.024 & 0.946 & & 0.984& & 0.987 & 0.959       &       & 255 \\ \midrule
    5     & 0.1   & 0.08  & 0.08  & 3     & 0.000  & 0.001 & 0.470 & 2.725 & -1.345 & -0.990 & -0.315 & 0.532 & 168891 \\
          & 0.2   & 0.13  & 0.13  & 3     & -0.001 & 0.002 & 0.297 & 4.390 & -2.169 & 0.674 & -0.323 & 0.556 & 10836 \\
          & 0.3   & 0.18  & 0.18  & 3     & 0.000 & 0.004 & 0.250 & 4.863 & -2.408 & 0.473 & -0.178 & 0.639 & 97784 \\
          & 0.4   & 0.26  & 0.26  & 3     & 0.001 & 0.004 & 0.228 & 2.778 & -1.371 & 0.513 & 0.043 & 0.726 & 7140 \\
          & 0.5   & 0.35  & 0.35  & 3     & 0.003 & 0.001 & 0.245 & 0.469 & -0.231 & 0.392 & 0.339 & 0.818 & 1450 \\
          & 0.6   & 0.46  & 0.46  & 3     & 0.001 & 0.005 & 0.293 & 2.113 & -1.047 & 0.642 & 0.598 & 0.886 & 3961 \\
          & 0.7   & 0.58  & 0.58  & 3     & 0.001 & 0.006 & 0.326 & 1.879 & -0.934 & 0.752 & 0.783 & 0.938 & 12785 \\
          & 0.8   & 0.71  & 0.71  & 2     & 0.016 & 0.038 &       & 0.860 &       & 0.896 & 0.949 &       & 392 \\
          & 0.9   & 0.85  & 0.85  & 2     & 0.015 & 0.035 &       & 0.970 &       & 0.967 & 0.987 &       & 394 \\
    \bottomrule
    \end{tabular}%
\end{table}

If two copula models are applied to discrete variables  and have the same strength of dependence as given by the Kendall's $\tau$, then the KL distance gets smaller. 
This is because the different asymptotic rates in the joint tails of the copula density do not affect rectangle probabilities for the log-likelihood with discrete response \citep{joe2014}. This means that a discretized  $K$-FNM copula model will be close to any copula model for discrete data even for strong dependence. 

To  show that we use   ordinal response variables, say $Y_1,Y_2$ with regressions on a scalar covariate $x$, which  is assumed to take $\mathcal{X}$ values equally spaced in $[-1, 1]$.  
Let $Z$ be a latent variable with cdf  $\mathcal{F}$,  such that $Y=y$ if
$\alpha_{y}+\beta x\leq Z\leq  \alpha_{y+1}+\beta x,\,y=0,\ldots,\mathcal{Y}-1,$
where $\mathcal{Y}$ is the number of categories of $Y$
(without loss of generality, we assume $\alpha_0=-\infty$ and  $\alpha_\mathcal{Y}=\infty$), and $\beta$ is the slope
of $x$. 
From this definition, the ordinal response $Y_j$ is assumed to have probability mass function (pmf)
$$\mathbb{P}(Y_j=y|x)=\mathcal{G}(\alpha_{y+1}+\beta_{j}x)-\mathcal{G}(\alpha_{y}+\beta_{j}x),\quad y=0,\ldots,\mathcal{Y}-1,\quad j=1,2.$$
Note that $\mathcal{G}$ normal leads to the probit model and $\mathcal{G}$ logistic
leads to the cumulative logit model for ordinal response. With copula families, the bivariate pmf (see e.g., \citealt{Nikoloulopoulos&karlis07MD})  can be obtained as 
\begin{eqnarray}\label{discrete-pmf}
f(y_1,y_2|x)&=&C\Bigl(\mathcal{G}(\alpha_{y_1+1}+\beta_{1}x),\mathcal{G}(\alpha_{y_2+1}+\beta_{2}x)\Bigr)-C\Bigl(\mathcal{G}(\alpha_{y_1}+\beta_{1}x),\mathcal{G}(\alpha_{y_2+1}+\beta_{2}x)\Bigr)-\nonumber\\&&C\Bigl(\mathcal{G}(\alpha_{y_1+1}+\beta_{1}x),G(\alpha_{y_2}+\beta_{2}x)\Bigr)+C\Bigl(\mathcal{G}(\alpha_{y_1}+\beta_{1}x),\mathcal{G}(\alpha_{y_2}+\beta_{2}x)\Bigr).
\end{eqnarray}
Let $f$ and $g$ denote the bivariate pmfs defined as in (\ref{discrete-pmf}) for the bivariate  Clayton and $K$-FNM  copula, respectively. Then the $\mbox{KL}(f,g)$ is
$$\mbox{KL}(f,g)=\mathlarger{\mathlarger{‎‎\sum}}_{x\in \mathcal{X}}\mathlarger{\mathlarger{‎‎\sum}}_{x\in \mathcal{X}}\mathlarger{\mathlarger{‎‎\sum}}_{y_1=0}^{\mathcal{Y}-1}\mathlarger{\mathlarger{‎‎\sum}}_{y_2=0}^{\mathcal{Y}-1} \log\left[\frac{f(y_1,y_2|x)}{g(y_1,y_2|x)}\right]f(y_1,y_2|x).$$

\setlength{\tabcolsep}{13pt}
\begin{table}[!h]
  \centering
  \caption{\label{Clayton-discrete}Minimized  KL distances, corresponding 2-FNM copula parameters and KL sample sizes for comparing the discretized Clayton copula model, as the Kendall's $\tau$ varies from 0.1 to 0.9,  versus the discretized  bivariate 2-FNM copula model. We use a varying number $\mathcal{Y}$ of ordinal categories (equally weighted)  from 2 to 5 and choose $\mathcal{X}=5$, $\beta_1=1$ and $\beta_2=0.7$. }
    \begin{tabular}{ccccccccc}
    \toprule
    $\tau$ & $\lambda_L$ & $\mathcal{Y}$ & $10^3\times \mbox{KL}(f,g)$ & $\pi_1$ & $\th_1$ & $\rho_1$ & $\rho_2$ & $n_{fg}$ \\
    \midrule
    0.1   & 0.04  & 2     & 0.003 & 0.950 & 0.908 & 0.080 & -0.143 & 1788690 \\
   
    0.2   & 0.25  &       & 0.007 & 0.885 & 0.890 & 0.150 & -0.898 & 721543 \\
    0.3   & 0.45  &       & 0.010 & 0.817 & 0.972 & 0.216 & -0.558 & 534811 \\
    0.4   & 0.59  &       & 0.015 & 0.744 & 0.989 & 0.289 & 0.187 & 353990 \\
    0.5   & 0.71  &       & 0.017 & 0.678 & 0.981 & 0.380 & 0.607 & 311204 \\
    0.6   & 0.79  &       & 0.006 & 0.609 & 0.987 & 0.486 & 0.822 & 838516 \\
    0.7   & 0.86  &       & 0.042 & 0.574 & 0.995 & 0.671 & 0.933 & 122823 \\
    0.8   & 0.92  &       & 0.091 & 0.559 & 0.992 & 0.854 & 0.981 & 55142 \\
    0.9   & 0.96  &       & 0.190 & 0.773 & 0.980 & 0.983 & 1.000 & 30454 \\\hline
    0.1   & 0.04  & 3     & 0.051 & 0.943 & 0.760 & 0.075 & -0.865 & 106069 \\
    0.2   & 0.25  &       & 0.139 & 0.884 & 0.863 & 0.145 & -0.430 & 39171 \\
    0.3   & 0.45  &       & 0.247 & 0.816 & 0.905 & 0.213 & 0.056 & 22213 \\
    0.4   & 0.59  &       & 0.331 & 0.746 & 0.916 & 0.284 & 0.470 & 16706 \\
    0.5   & 0.71  &       & 0.248 & 0.674 & 0.944 & 0.358 & 0.720 & 22017 \\
    0.6   & 0.79  &       & 0.196 & 0.609 & 0.990 & 0.467 & 0.856 & 27265 \\
    0.7   & 0.86  &       & 0.503 & 0.555 & 1.022 & 0.625 & 0.938 & 9672 \\
    0.8   & 0.92  &       & 0.791 & 0.503 & 1.022 & 0.802 & 0.981 & 5770 \\
    0.9   & 0.96  &       & 0.690 & 0.554 & 1.018 & 0.959 & 1.000 & 8247 \\\hline
    0.1   & 0.04  & 4     & 0.114 & 0.945 & 0.755 & 0.075 & -0.608 & 47632 \\
    0.2   & 0.25  &       & 0.330 & 0.884 & 0.832 & 0.144 & -0.218 & 16580 \\
    0.3   & 0.45  &       & 0.598 & 0.820 & 0.875 & 0.214 & 0.206 & 9236 \\
    0.4   & 0.59  &       & 0.774 & 0.754 & 0.905 & 0.286 & 0.545 & 7137 \\
    0.5   & 0.71  &       & 0.704 & 0.683 & 0.948 & 0.363 & 0.743 & 7673 \\
    0.6   & 0.79  &       & 0.775 & 0.611 & 0.993 & 0.464 & 0.859 & 6808 \\
    0.7   & 0.86  &       & 1.429 & 0.546 & 1.024 & 0.609 & 0.935 & 3410 \\
    0.8   & 0.92  &       & 2.070 & 0.486 & 1.028 & 0.781 & 0.980 & 2213 \\
    0.9   & 0.96  &       & 2.387 & 0.622 & 1.024 & 0.964 & 1.000 & 2833 \\\hline
    0.1   & 0.04  & 5     & 0.175 & 0.946 & 0.748 & 0.075 & -0.511 & 31010 \\
    0.2   & 0.25  &       & 0.525 & 0.886 & 0.817 & 0.144 & -0.121 & 10462 \\
    0.3   & 0.45  &       & 0.956 & 0.823 & 0.861 & 0.214 & 0.280 & 5787 \\
    0.4   & 0.59  &       & 1.216 & 0.757 & 0.899 & 0.285 & 0.582 & 4518 \\
    0.5   & 0.71  &       & 1.186 & 0.687 & 0.948 & 0.362 & 0.757 & 4535 \\
    0.6   & 0.79  &       & 1.449 & 0.615 & 0.996 & 0.464 & 0.864 & 3616 \\
    0.7   & 0.86  &       & 2.583 & 0.549 & 1.027 & 0.606 & 0.936 & 1892 \\
    0.8   & 0.92  &       & 3.752 & 0.482 & 1.029 & 0.772 & 0.979 & 1229 \\
    0.9   & 0.96  &       & 3.729 & 0.490 & 1.020 & 0.941 & 1.000 & 1593 \\
    \bottomrule
    \end{tabular}%
\end{table}

Table \ref{Clayton-discrete} shows the minimized  KL distances, the corresponding 2-FNM copula parameters and KL sample sizes for comparing the discretized Clayton  copula model, as the Kendall's $\tau$ varies from 0.1 to 0.9,  versus the discretized   2-FNM copula model. We show the comparison results versus the Clayton copula, as  in Table \ref{1-par-comparison} it
was revealed that the Clayton copula is the 1-parameter copula family which is the most far apart from the 2-FNM copula for continuous responses.
We used univariate ordinal regressions, but note that using ordinal probit regressions led to similar results. 

 The conclusions from the table and the  other computations we have done  for other copula families are:

\begin{itemize}
\itemsep=10pt
\item The $K$-FNM copula  is close to any parametric bivariate family of copulas if two copulas models are applied to discrete variables. 
\item With discrete response variables, it takes larger sample sizes to distinguish the $K$-FNM copula (because tails of the copula densities would not be ``observed"). 
\item The KL distances (sample sizes) get larger (smaller) with less discretization, i.e., as $\mathcal{Y}$ increases. 
\end{itemize} 

\section{\label{simulations}Simulations}
To gauge the small sample efficiency  of the ML estimation method in Section \ref{MLsection}
to estimate the $K$-FNM copula parameters, we performed several simulation studies using $K$-FNM copula
models with various parameter choices for $K=2,3$. We report
here typical results from these experiments.

We randomly generated $B=10^4$ samples of size {\bf $n = 100, 300, 500$} from the 2-and 3-FNM bivariate copulas with exponential margins  and  marginal parameters $\lambda_1=0.5$ and $\lambda_2=1$. We have transformed the simulated data  to uniform random variables   using  their empirical   distributions, i.e., we have approached estimation of the $K$-FNM copula parameters  using the semi-parametric estimation \citep{gen&ghoudi&riv95}. We have used as initial values  the ones  that resemble the independence copula.

Table~\ref{biasmse} contains the copula parameter values, the bias,
standard deviation (SD) and the root mean square errors (RMSE) of the
ML estimates,  along
with the  average of their theoretical SDs.
The theoretical SD of the ML estimate is obtained
via the gradients and the Hessian computed
numerically during the maximization process. The conclusions from the table and the  other computations we have done are that
\begin{itemize}
\item ML  is highly efficient according to the simulated biases, SDs and RMSEs as the sample size increases. 
\item The  SDs computed from the simulations are close to the asymptotic SDs as the sample size increases.
\item For small samples the estimates of the  mean parameters $(\th_1,\ldots,\th_{K-1})$ have upward bias.   
\end{itemize}

\begin{table}[!h]
 \centering
 \caption{\label{biasmse}Small sample of sizes $N = 100, 300, 500$ simulations ($10^4$ replications) from the 2- and 3-FNM copula model  with exponential  margins and biases, root mean square errors (RMSEs) and standard deviations (SDs), along with the square root of the average theoretical variances $(\sqrt{V})$ for the MLEs.}
  \setlength{\tabcolsep}{25pt}
   \begin{tabular}{cccccc}
  
   \toprule
   \multicolumn{6}{c}{$K=2$} \\
   
         & $n$ & $\pi_1=0.3$ & $\theta_1=0$ & $\rho_1=0.8$ & $\rho_2=-0.8$ \\
         \midrule
   Bias  & 100   & 0.001 & 0.017 & 0.001 & -0.001 \\
         & 300   & -0.001 & 0.009 & 0.002 & -0.001 \\
         & 500   & -0.001 & 0.005 & 0.001 & -0.001 \\
   SD    & 100   & 0.074 & 0.170 & 0.090 & 0.048 \\
         & 300   & 0.039 & 0.086 & 0.042 & 0.026 \\
         & 500   & 0.030 & 0.066 & 0.032 & 0.020 \\
   $\sqrt{V}$ & 100   & 0.046 & 0.111 & 0.066 & 0.039 \\
         & 300   & 0.026 & 0.058 & 0.037 & 0.022 \\
         & 500   & 0.019 & 0.042 & 0.028 & 0.017 \\
   RMSE  & 100   & 0.074 & 0.171 & 0.090 & 0.048 \\
         & 300   & 0.039 & 0.087 & 0.042 & 0.026 \\
         & 500   & 0.030 & 0.066 & 0.032 & 0.020 \\
   
   \end{tabular}%
  \setlength{\tabcolsep}{9pt}
   \begin{tabular}{ccccccccc}
   \toprule
   \multicolumn{9}{c}{$K=3$} \\
         & $n$ & $\pi_1=0.2$ & $\pi_1=0.3$ & $\theta_1=0.5$ & $\theta_2=0.5$ & $\rho_1=0.8$ & $\rho_2=-0.8$ & $\rho_3=0.8$ \\
   \midrule
   Bias  & 100   & -0.028 & -0.018 & 0.704 & 0.258 & -0.093 & 0.088 & -0.069 \\
         & 300   & -0.002 & -0.001 & 0.113 & 0.041 & -0.009 & 0.006 & -0.005 \\
         & 500   & -0.001 & 0.000 & 0.031 & -0.001 & 0.000 & 0.000 & 0.001 \\
   SD    & 100   & 0.120 & 0.091 & 1.054 & 0.932 & 0.320 & 0.222 & 0.205 \\
         & 300   & 0.061 & 0.042 & 0.438 & 0.360 & 0.088 & 0.066 & 0.056 \\
         & 500   & 0.041 & 0.031 & 0.209 & 0.182 & 0.064 & 0.038 & 0.030 \\
   $\sqrt{V}$ & 100   & 0.028 & 0.034 & 0.290 & 0.324 & 0.252 & 0.116 & 0.061 \\
         & 300   & 0.022 & 0.024 & 0.089 & 0.086 & 0.060 & 0.041 & 0.031 \\
         & 500   & 0.019 & 0.019 & 0.064 & 0.063 & 0.037 & 0.031 & 0.023 \\
   RMSE  & 100   & 0.123 & 0.093 & 1.268 & 0.967 & 0.333 & 0.238 & 0.217 \\
         & 300   & 0.061 & 0.042 & 0.453 & 0.363 & 0.089 & 0.066 & 0.057 \\
         & 500   & 0.041 & 0.031 & 0.211 & 0.182 & 0.064 & 0.038 & 0.030 \\
   \bottomrule
   \end{tabular}%
\end{table}

\section{\label{Appsection}Empirical examples}
In this section we illustrate the proposed methodology by analysing two  real data examples with distinct dependence structures,   the first in the area of astrophysics and the second in agriculture. In Section \ref{magic-sec} we analyse the  two aforementioned MAGIC variables  with peculiar dependence that  typical bivariate copulas fail to model, while in Section \ref{nut-sec} we analyse three variables from the 1985  survey of nutritional habits commissioned  by the United States Department of Agriculture that have  strong reflection asymmetric dependence. 

We estimate each marginal distribution non-parametrically by the empirical distribution function of $Y_j$, viz.  
$$u_{ij}=F_j(y_{ij})=\frac{1}{n+1}\sum_{i=1}^n\mathbf{1}(Y_{ij}\leq y_{ij})=r_{ij}/(n+1),$$
where $r_{ij}$ denotes the rank of
$y_{ij}$. Hence we  allow the distribution of the continuous margins to be quite free and not restricted by parametric families. 
We use simple diagnostics   to identify the suitable  copula family. Although copula theory uses transforms to standard uniform margins, for diagnostics, we convert  the original data  to normal scores using the normal quantiles of their empirical distributions.  With a bivariate normal scores plot  \citep{nikoloulopoulos&joe&li11} one can check for deviations from the elliptical shape that would be expected with the BVN copula, and hence assess if tail asymmetry  exists on the data.

Having discussed why more flexible dependencies are needed
we proceed with the $K$-FNM copula models and construct a plausible $K$-FNM copula model, to capture any type of  dependence. For a baseline comparison, we  initially  fit the typical copula families presented in Section \ref{bivcop}. 
 To make it easier to compare strengths of dependence, we convert the estimated copula parameters to Kendall's $\tau$'s, lower tail dependence $\lambda_L$ and upper tail dependence $\lambda_U$ via the relations in \citet[Chapter 4]{joe2014}. The  estimated Kendall's $\tau$ of the $K$-FNM copula, viz.  

\begin{small}
\begin{multline*}
\hat\tau(Y_1,Y_2;\hat\pi_1,\ldots,\hat\pi_{K-1},\hat\th_1,\ldots,\hat\th_{K-1},\hat\rho_1,\ldots,\hat\rho_{K}) =\\
 -1 + 4\int_0^1\int_0^1C(u_1,u_2;\hat\pi_1,\ldots,\hat\pi_{K-1},\hat\th_1,\ldots,\hat\th_{K-1},\hat\rho_1,\ldots,\hat\rho_{K})dC(u_1,u_2;\hat\pi_1,\ldots,\hat\pi_{K-1},\hat\th_1,\ldots,\hat\th_{K-1},\hat\rho_1,\ldots,\hat\rho_{K}),
\end{multline*}
\end{small}

\noindent has been calculated via adaptive bivariate integration over hypercubes \citep{Narasimhan-etal-2018};  $C(\cdot;\cdot)$ is the $K$-FNM copula cdf given in (\ref{FNMcdf}).

To find a copula model that provides a good fit   we don't use goodness-of-fit procedures (see e.g., \citealt{genest&Remillard&Beaudoin09} and the references therein), but we rather  adopt the Akaike's information criterion (AIC).  The goodness-of-fit procedures involve a global distance measure between the model-based and empirical distribution, hence they might not be sensitive to tail behaviours and are not diagnostic in the sense of suggesting improved parametric models in the case of small $p$-values \citep{joe2014}.  For vine copulas, \cite{Dissmann-etal-2013-csda} found that pair-copula selection based on likelihood and AIC seem to be better than using bivariate goodness-of-fit tests.  The AIC is  $$-2\times \ell+2\times \mbox{(\# model parameters) }$$ and a smaller AIC value indicates a copula model better approximates both  the dependence structure of the data, and the strength of dependence in the tails.

\subsection{\label{magic-sec}MAGIC  telescope}

Ground-based atmospheric Cherenkov telescopes using the imaging technique are a useful  addition to the variety of instruments used by astrophysicists. The MAGIC telescope, located on the Canary islands, observes high-energy gamma rays,  detecting the radiation emitted by charged particles produced inside electromagnetic showers. Depending on the energy of the primary gamma,   Cherenkov photons get collected, in patterns (called the shower image), allowing to discriminate statistically those caused by primary gammas (signal) from the images of hadronic showers initiated by cosmic rays in the upper atmosphere (background). Typically, the image of a shower is an elongated cluster; its long axis is oriented towards the camera center if the shower axis is parallel to the telescope's optical axis, i.e. if the telescope axis is directed towards a point source. If the depositions were distributed as a BVN, this would be an equidensity ellipse. The characteristic parameters of this ellipse  are among the image parameters. The energy depositions are typically asymmetric along the major axis \citep{bock-etal-2004}. 

\begin{figure}[!h]

\begin{center}
\caption{\label{nscores-magic}   Bivariate normal scores plot for  the Length  and M3Long variables. }
\vspace{-2ex}
\includegraphics[width=0.7\textwidth]{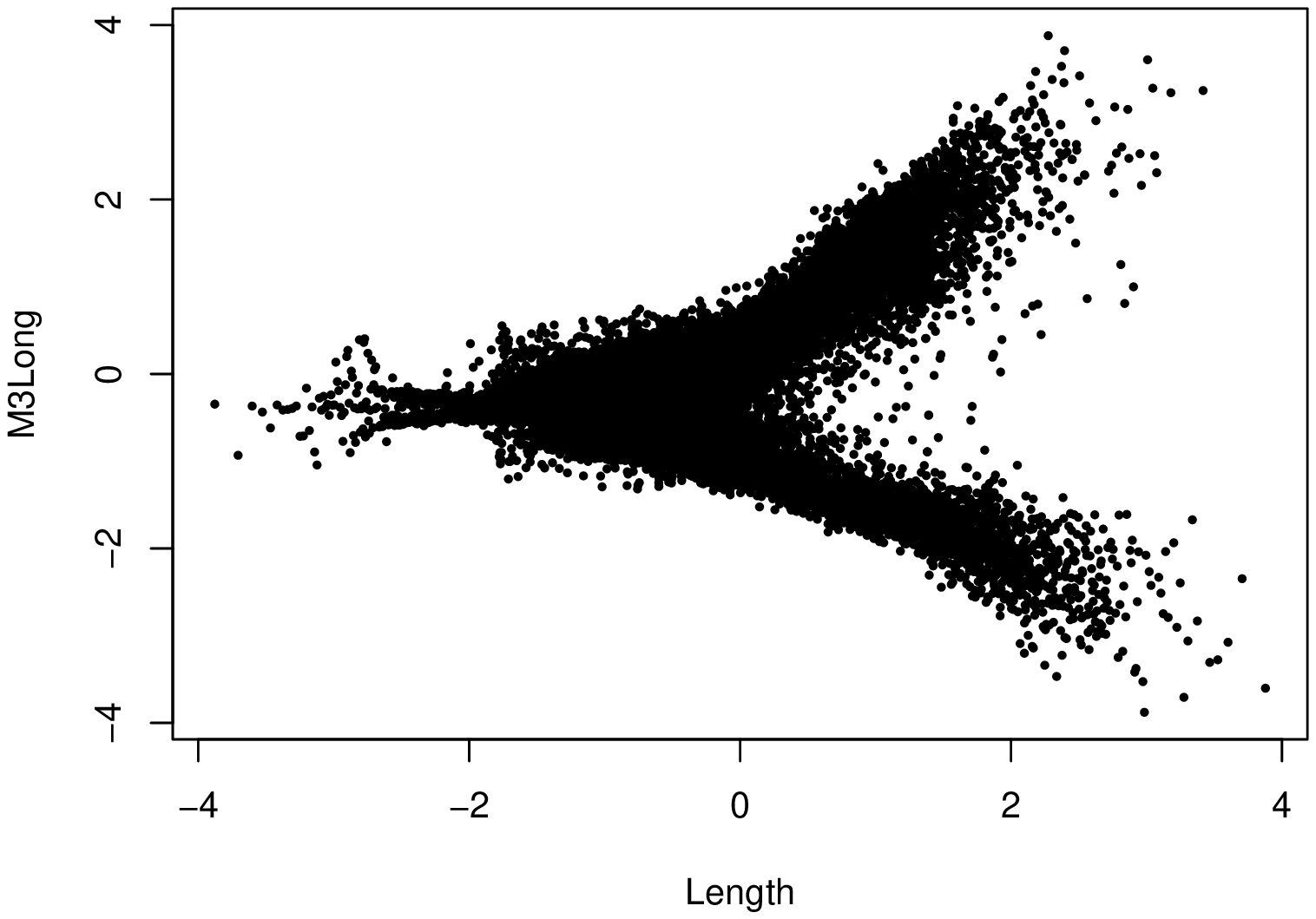}
\end{center}
\end{figure}

We apply the $K$-FNM copula to  2 out of 10  MAGIC image parameters in \cite{bock-etal-2004}. Our objective is to describe the joint distribution of the Length  and M3Long  that have a peculiar dependence.  
The data set with the 10 MAGIC image parameters is available from the \href{https://archive.ics.uci.edu/ml/datasets/magic+gamma+telescope}{UCI Machine Learning Repository web page} and  comprises $n=19,020$ observations.      In Figure \ref{nscores-magic} we depict the bivariate normal scores plot for the Length  and M3Long. 
From the plot, it  is revealed  that none of the existing parametric families of copulas can adequately  model  the dependence structure between the variables.

Table \ref{other-copula-fit-magic} gives the AICs,  estimated copula parameters and their SE, along with the family-based  Kendall's $\tau$ and tail dependence parameters $\lambda_L,\lambda_U$ for each fitted parametric family of copulas.  The AICs show, that among the existing parametric families of copulas,  the $t_\delta$ copula  provides the best  fit.

\begin{table}[!h]
 \centering
 \caption{\label{other-copula-fit-magic}AICs,  estimated copula parameters and their standard errors (SE), along with the model-based  Kendall's $\tau$ and tail dependence parameters $\lambda_L,\lambda_U$ for each fitted parametric family of copulas for  the Length  and M3Long variables. }
 \setlength{\tabcolsep}{11pt}
   \begin{tabular}{lcccccccc}
   \toprule
   Copula & AIC   & $\theta$ & SE    & $\delta$ & SE    & $\tau$ & $\lambda_L$ & $\lambda_U$ \\
   \midrule
   BVN   & -648.1 & 0.183 & 0.007 &       &       & 0.117 &  &  \\
   $t_\delta$ & -4590.3 & 0.352 & 0.008 & 2.159 & 0.042 & 0.229 & 0.302 & 0.302 \\
   Clayton & 2.1 & 0.000  & 0.002 &       &      & 0.000  & 0.000 &  \\
   Gumbel & -3069.4 & 1.314 & 0.007 &       &       & 0.239 &  & 0.305 \\
   Frank & -2004.5 & 2.167 & 0.049 &       &       & 0.230 &  &  \\
   BB1   & -3059.3 & 0.001 &  & 1.314 & 0.007 & 0.239 &  & 0.305 \\
   BB7   & -4110.6 & 1.591 & 0.013 & 0.001 & 0.021 & 0.249 &  & 0.454 \\
   Survival Clayton & -3363.7 & 0.651 & 0.013 &       &       & 0.246 &  & 0.345 \\
   Survival Gumbel & -228.1 & 1.102 & 0.007 &       &       & 0.093 & 0.125 &  \\
   Survival BB1 & -3353.45 & 0.650 & 0.022 & 1.001 & 0.012 & 0.246 & 0.001 & 0.500 \\
   Survival BB7 & -3355.2 & 1.001 & 0.013 & 0.651 & 0.014 & 0.246 & 0.001 & 0.345 \\
   \bottomrule
   \end{tabular}%
\end{table}

\begin{table}[!h]
 \centering
 \caption{\label{fnm-magic}AICs,  estimated $K$-FNM copula parameters and their standard errors (SE), along with the family-based  Kendall's $\tau$ for different number of components $K$ for  the Length  and M3Long variables.}
 \setlength{\tabcolsep}{25pt}
   \begin{tabular}{ccccccc}
   \toprule
         &       & \multicolumn{2}{c}{$K=2$} &       & \multicolumn{2}{c}{$K=3$} \\
   
          &       & Est.  & SE    &       & Est.  & SE \\         
\cmidrule{1-1}   \cmidrule{3-4} \cmidrule{6-7}
   $\pi_1$ &       & 0.127 & 0.001 &       & 0.001 & 0.000 \\
   $\pi_2$ &       &       &       &       & 0.334 & 0.000  \\
   $\th_1$ &       & -1.882 & 0.016 &       & -1.045 & 0.002 \\
   $\th_2$ &       &       &       &       & -1.145 & 0.001 \\
   $\rho_1$ &       & -0.784 & 0.006 &       & -0.470 & 0.103 \\
   $\rho_2$ &       & 0.747 & 0.003 &       & -0.854 & 0.003 \\
   $\rho_3$ &       &       &       &       & 0.901 & 0.001 \\\cmidrule{1-1}   \cmidrule{3-4} \cmidrule{6-7}
   $\tau$ &       & \multicolumn{2}{c}{0.304} &       & \multicolumn{2}{c}{0.310} \\
   AIC   &       & \multicolumn{2}{c}{-17320.5} &       & \multicolumn{2}{c}{-27064.1} \\

   \bottomrule
   \end{tabular}%
\end{table}

\begin{figure}[!h]
\begin{center}
\caption{\label{figure-fit-magic} Estimated contour plots of   the 2- and 3-FNM copulas with standard normal margins, along with the bivariate normal scores plot  for  the Length  and M3Long variables.}
\begin{tabular}{|cc|}
\hline
$K=2$&$K=3$\\\hline
\includegraphics[width=0.45\textwidth]{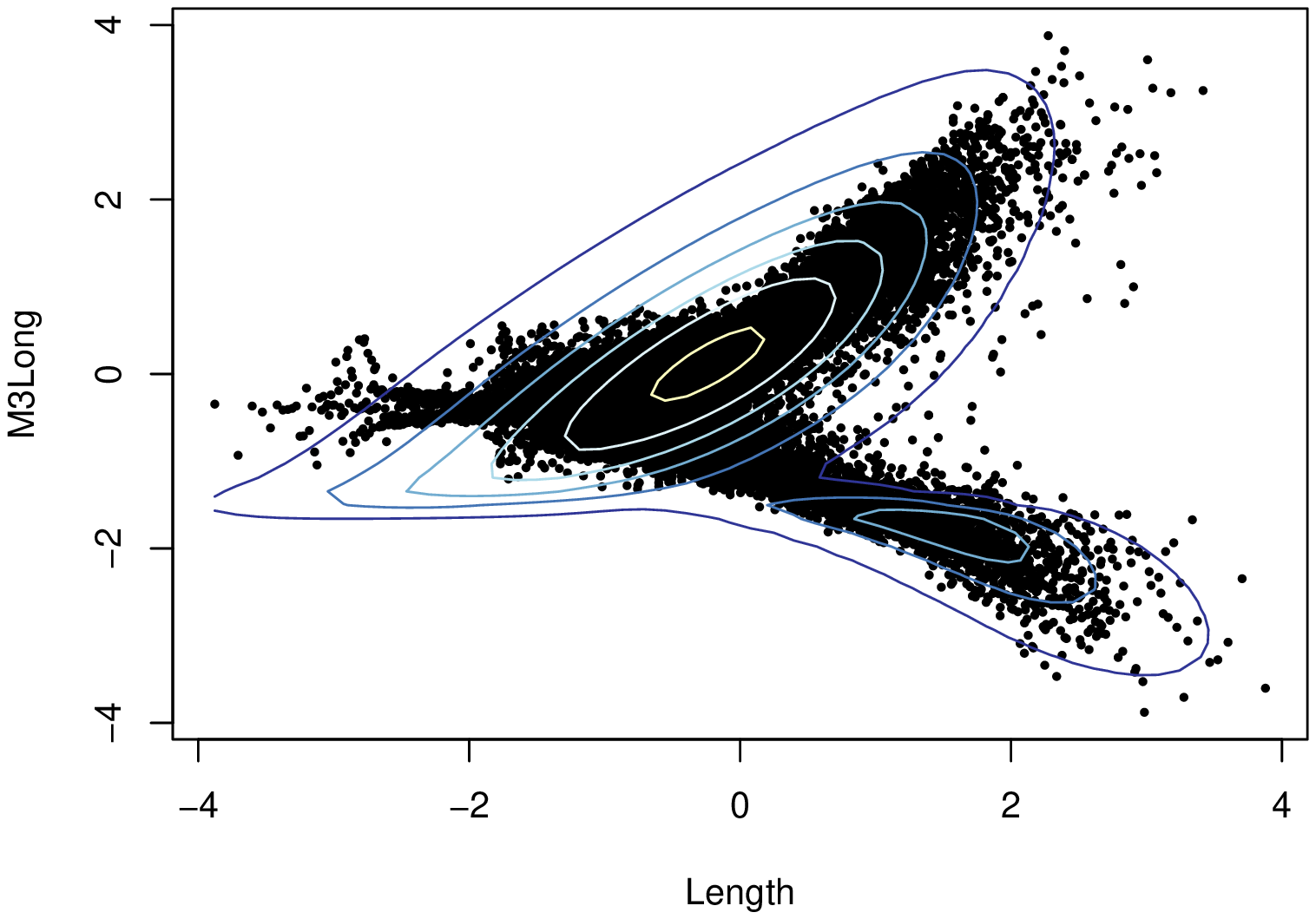}
&
\includegraphics[width=0.45\textwidth]{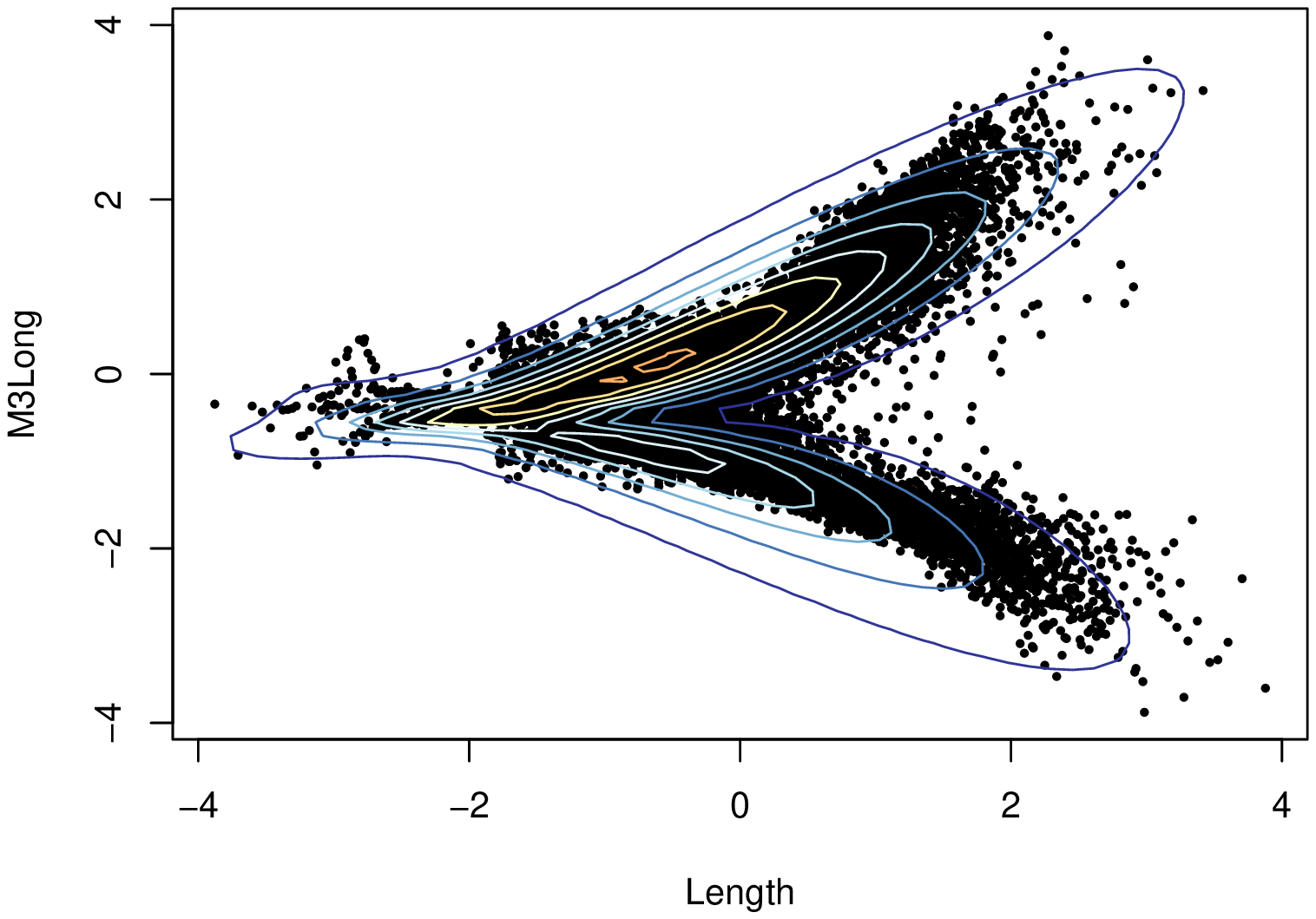}\\\hline
\end{tabular}
\end{center}
\end{figure}

Then we exploit the use of  the $K$-FNM copula  
to construct a plausible copula family to represent the joint distribution of  Length  and M3Long. Table \ref{fnm-magic} gives the AICs,  estimated copula parameters and their SE, along with the family-based  Kendall's $\tau$ for different numbers of components.   
The AICs show, that  the 3-FNM copula provides the best fit and provides much better fit than the $t_\delta$, since the AIC has been improved by $22473.8=-4590.3-(-27064.1)$.  
Note in passing that using $K>3$,  the estimated mixing probabilities for the extra components were close to zero, and, hence, there was no improvement in fit. 
In Figure \ref{figure-fit-magic} we depict the estimated contour plots of the 2- and 3-FNM copulas with standard normal margins, along with the bivariate normal scores plots.
From the plots, it  is revealed  that the 3-FNM copula provides a realistic representation of the joint distribution.

The new-parametric family of copulas does not only allow to make accurate inferences that are based on the joint distribution, but also  provides superior statistical inference for the parameters of interest, such as Kendall's $\tau$. From Table \ref{FNM-fit}, it is revealed that the Kendall's $\tau$  was underestimated using simple parametric families of copulas and a change from a $\tau$-value  of 0.23 ($t_\delta$-based) to one slightly larger than 0.30 has been achieved. 

\subsection{\label{nut-sec}Nutritional habits}
\begin{figure}[!h]
\caption{\label{normal-scores}  Bivariate normal scores plots for the pairs identified as asymmetric  in the nutrient data set. } 

\vspace{-0.5cm}

\begin{center}
\begin{tabular}{|cc|}
\hline
\includegraphics[width=0.45\textwidth]{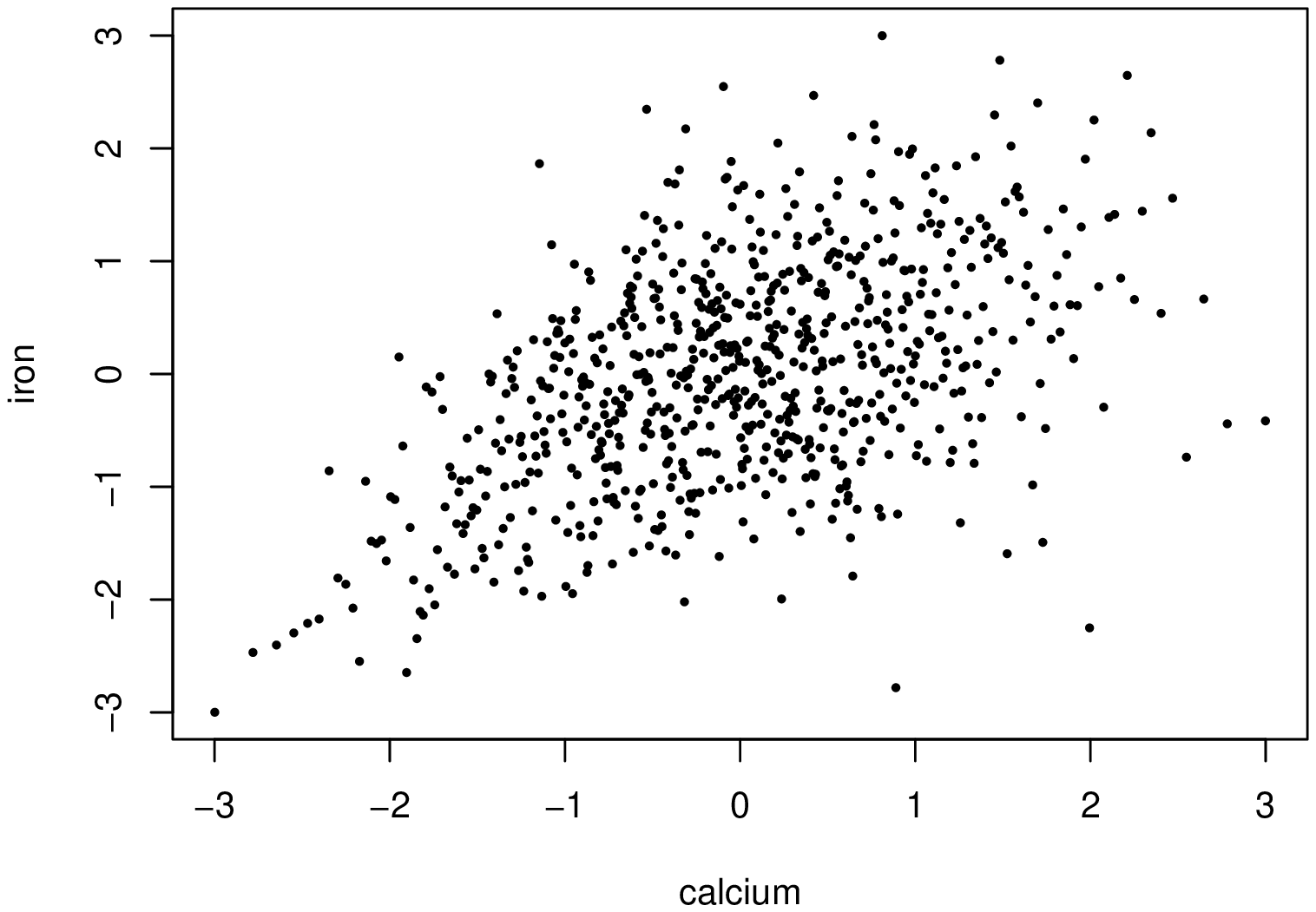}&
\includegraphics[width=0.45\textwidth]{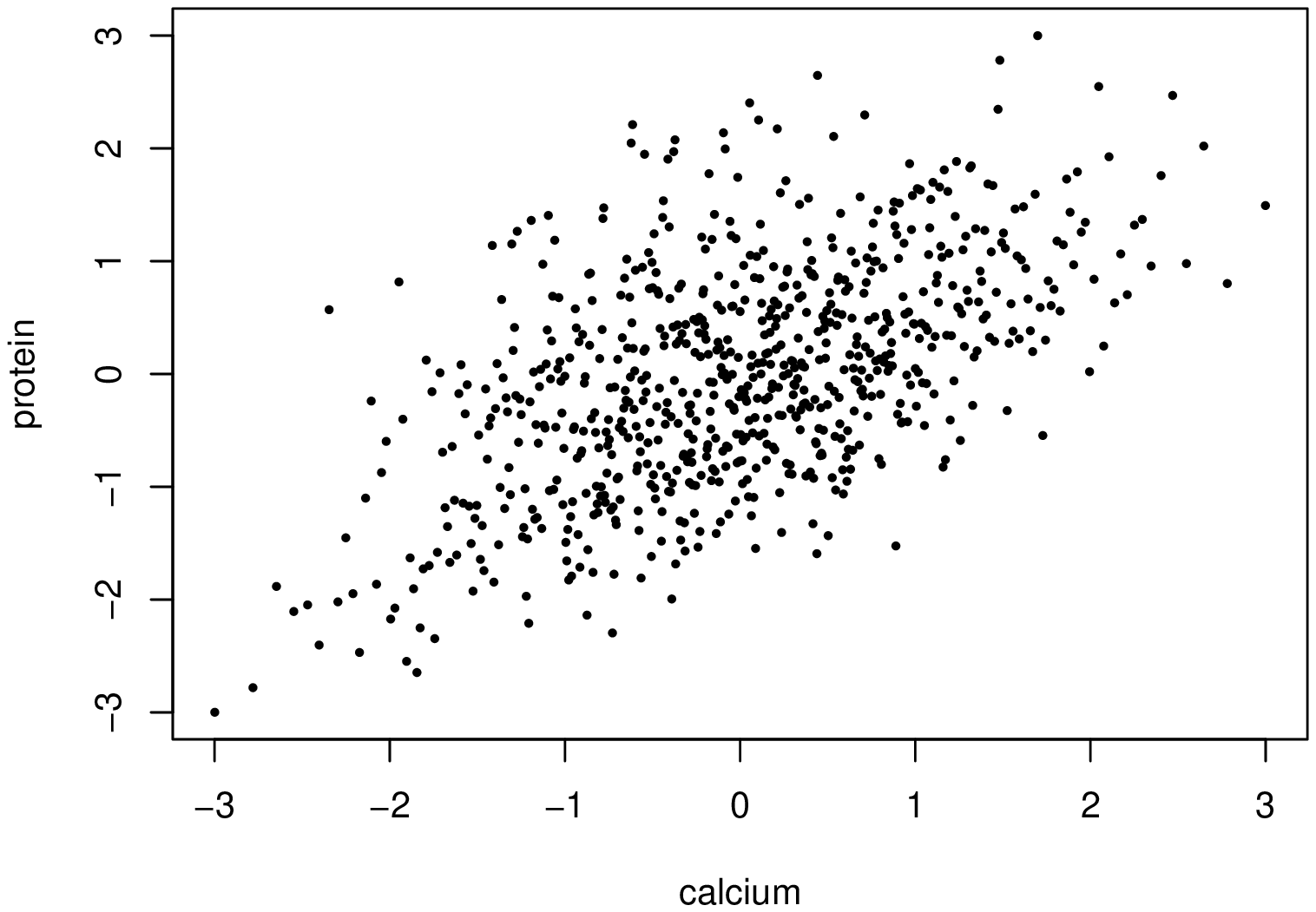}\\
\hline
\end{tabular}

\end{center}
\end{figure}

\cite{McNeil&Neslehova-2010-jmva} analysed three variables, namely daily calcium intake (in mg), daily iron intake (in mg) and daily protein intake (in g), 
of $n$=747 female respondents  aged between 25 and 50 years to the 1985  survey of nutritional habits commissioned  by the United States Department of Agriculture.
This dataset  and its description  can be found in the {\tt R} package {\tt lcopula} \citep{belzile-etal-2019}.  \cite{genest-etal-2012} identified  a strongly asymmetric dependence structure between the intakes of calcium and iron, and between the intakes of calcium and protein. 
In this section we apply the $K$-FNM copula to   the pairs identified as asymmetric to illustrate that it can sufficiently allow for reflection asymmetric dependence.
In Figure \ref{normal-scores} we depict the bivariate normal scores plots for the pairs identified as asymmetric.
From the plots, it  is revealed that there is more skewness in the lower tail.

\begin{table}[!h]
  \centering
  \caption{\label{other-copula-fit}AICs,  estimated copula parameters and their standard errors (SE), along with the model-based  Kendall's $\tau$ and tail dependence parameters $\lambda_L,\lambda_U$ for each fitted parametric family of copulas  for the pairs identified as asymmetric  in the nutrient data set.}
  \setlength{\tabcolsep}{11pt}
    \begin{tabular}{lcccccccc}
    \toprule
   \multicolumn{9}{l}{ \underline{Calcium and Iron}}\\
    Copula & AIC & $\hat\theta$ & SE    & $\hat\delta$ & SE    & $\hat\tau$ & $\hat\lambda_L$ & $\hat\lambda_U$ \\
  \hline
 BVN   & -203.0 & 0.497 & 0.025 &  &  & 0.331 &  &  \\
   \midrule
   $t_\delta$ & -216.6 & 0.492 & 0.030 & 6.563 & 2.026 & 0.328 & 0.149 & 0.149 \\
   Clayton & -230.7 & 0.885 & 0.069 &  &  & 0.307 & 0.457 &  \\
   Gumbel & -162.0 & 1.412 & 0.040 &  &  & 0.292 &  & 0.366 \\
   Frank & -173.0 & 3.140 & 0.238 &  &  & 0.319 &  &  \\
   BB1   & -238.3 & 0.684 & 0.091 & 1.115 & 0.043 & 0.332 & 0.403 & 0.138 \\
   BB7   & -238.9 & 1.165 & 0.059 & 0.807 & 0.075 & 0.329 & 0.424 & 0.187 \\
   Survival Clayton & -114.8 & 0.582 & 0.061 &  &  & 0.225 &  & 0.304 \\
   Survival Gumbel & -239.6 & 1.490 & 0.043 &  &  & 0.329 & 0.408 &  \\
   Survival BB1 & -237.7 & 0.016 & 0.063 & 1.480 & 0.057 & 0.330 & 0.403 & 0.626 \\
   Survival BB7 & -240.6 & 1.611 & 0.069 & 0.270 & 0.070 & 0.320 & 0.462 & 0.077 \\
\hline
 \multicolumn{9}{l}{ \underline{Calcium and Protein}}\\
  Copula & AIC & $\hat\theta$ & SE    & $\hat\delta$ & SE    & $\hat\tau$ & $\hat\lambda_L$ & $\hat\lambda_U$ \\
    \hline
 BVN   & -267.8 & 0.558 & 0.022 &  &  & 0.377 &  &  \\
   $t_\delta$ & -268.9 & 0.553 & 0.025 & 12.323 & 6.752 & 0.373 & 0.072 & 0.072 \\
   Clayton & -261.7 & 0.965 & 0.071 &  &  & 0.325 & 0.487 &  \\
   Gumbel & -217.2 & 1.499 & 0.043 &  &  & 0.333 &  & 0.412 \\
   Frank & -227.2 & 3.657 & 0.244 &  &  & 0.362 &  &  \\
   BB1   & -282.3 & 0.633 & 0.091 & 1.196 & 0.049 & 0.365 & 0.401 & 0.215 \\
   BB7   & -281.3 & 1.264 & 0.066 & 0.838 & 0.079 & 0.357 & 0.437 & 0.270 \\
   Survival Clayton & -166.0 & 0.714 & 0.064 &  &  & 0.263 &  & 0.379 \\
   Survival Gumbel & -283.3 & 1.567 & 0.045 &  &  & 0.362 & 0.444 &  \\
   Survival BB1 & -284.4 & 0.115 & 0.068 & 1.493 & 0.060 & 0.367 & 0.409 & 0.629 \\
   Survival BB7 & -284.6 & 1.632 & 0.073 & 0.407 & 0.074 & 0.354 & 0.471 & 0.182 \\
    \bottomrule
    \end{tabular}%
\end{table}

Table \ref{other-copula-fit} gives the AICs,  estimated copula parameters and their SE, along with the family-based  Kendall's $\tau$ and tail dependence parameters $\lambda_L,\lambda_U$ for each fitted parametric family of copulas.  The AICs show, that among the existing parametric families of copulas,  the survival BB7 copula  provides the best  fit for both pairs identified as asymmetric.

\begin{table}[!h]
  \centering
  \caption{\label{FNM-fit}AICs and estimated $2$-FNM copula parameters along with their standard errors (SE) for the pairs identified as asymmetric  in the nutrient data set.}
  \setlength{\tabcolsep}{25pt}
  \begin{tabular}{ccccccc}
    \toprule
    
    &&\multicolumn{2}{c}{{Calcium and Iron}}&&\multicolumn{2}{c}{{Calcium and Protein}}\\
   
          &       & Est.  & SE    &       & Est.  & SE \\
\cmidrule{1-1}    \cmidrule{3-4} \cmidrule{6-7}
$\pi$ &       & 0.848 & 0.055 &       & 0.953 & 0.008 \\
   $\theta$ &       & 0.518 & 0.136 &       & 2.012 & 0.094 \\
   $\rho_1$ &       & 0.339 & 0.044 &       & 0.474 & 0.030 \\
   $\rho_2$ &       & 0.779 & 0.062 &       & 0.594 & 0.108 \\
   \cmidrule{1-1}    \cmidrule{3-4} \cmidrule{6-7}
   $\tau$ &       & \multicolumn{2}{c}{0.330} &       & \multicolumn{2}{c}{0.341} \\
   AIC   &       & \multicolumn{2}{c}{-243.7} &       & \multicolumn{2}{c}{-291.7} \\
    \bottomrule
    \end{tabular}%
\end{table}

\begin{figure}[!h]
\caption{\label{app-contours} estimated contour plots of the 2-FNM and survival BB7 copulas  with standard normal margins, along with the bivariate normal scores plot for the pairs identified as asymmetric  in the nutrient data set. } 

\vspace{-0.5cm}

\begin{center}
\begin{tabular}{|cc|}
\hline
2-FNM &  Survival BB7 \\\hline
\includegraphics[width=0.45\textwidth]{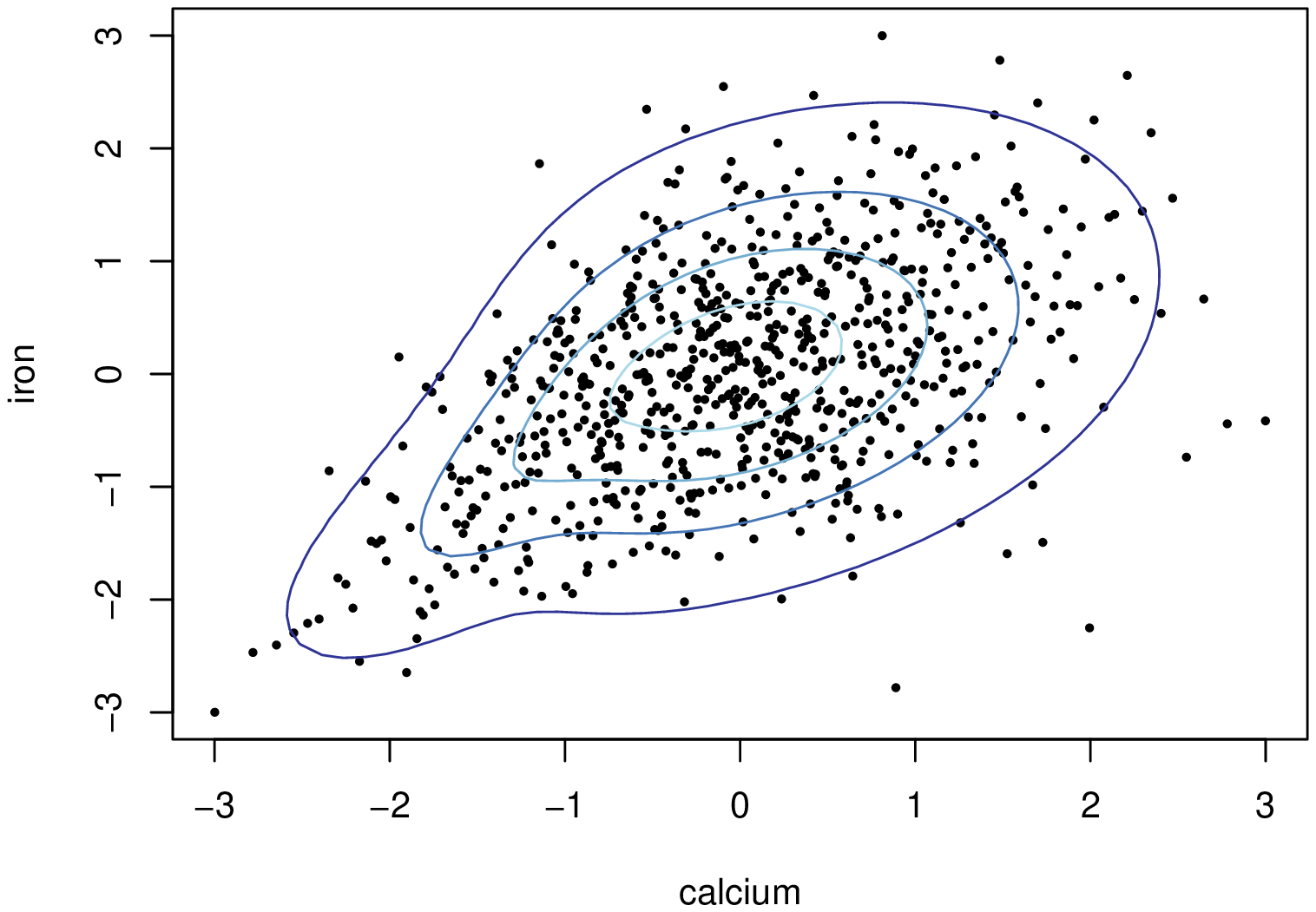}&
\includegraphics[width=0.45\textwidth]{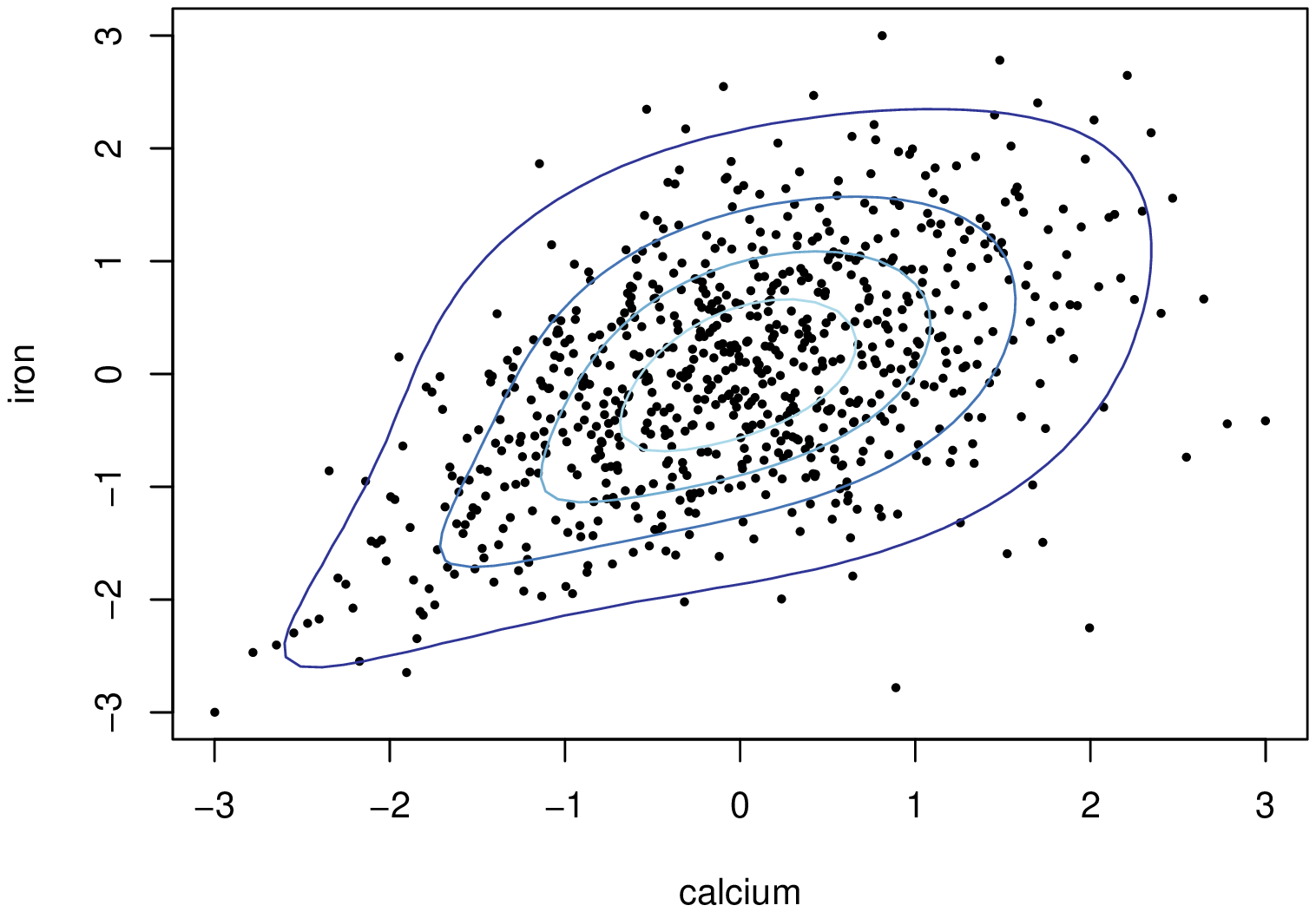} \\\hline
2-FNM &  Survival BB7 \\\hline

\includegraphics[width=0.45\textwidth]{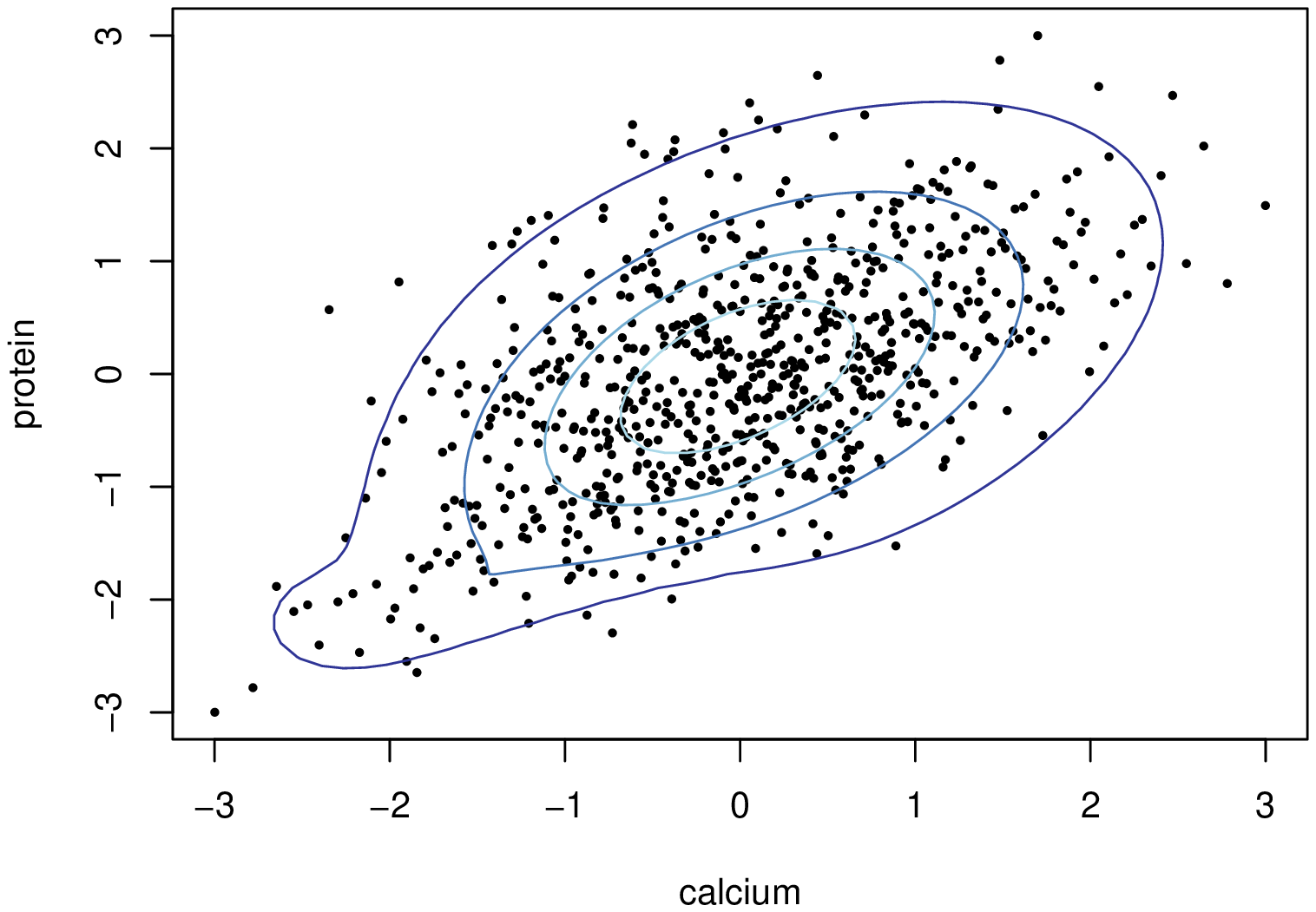} &\includegraphics[width=0.45\textwidth]{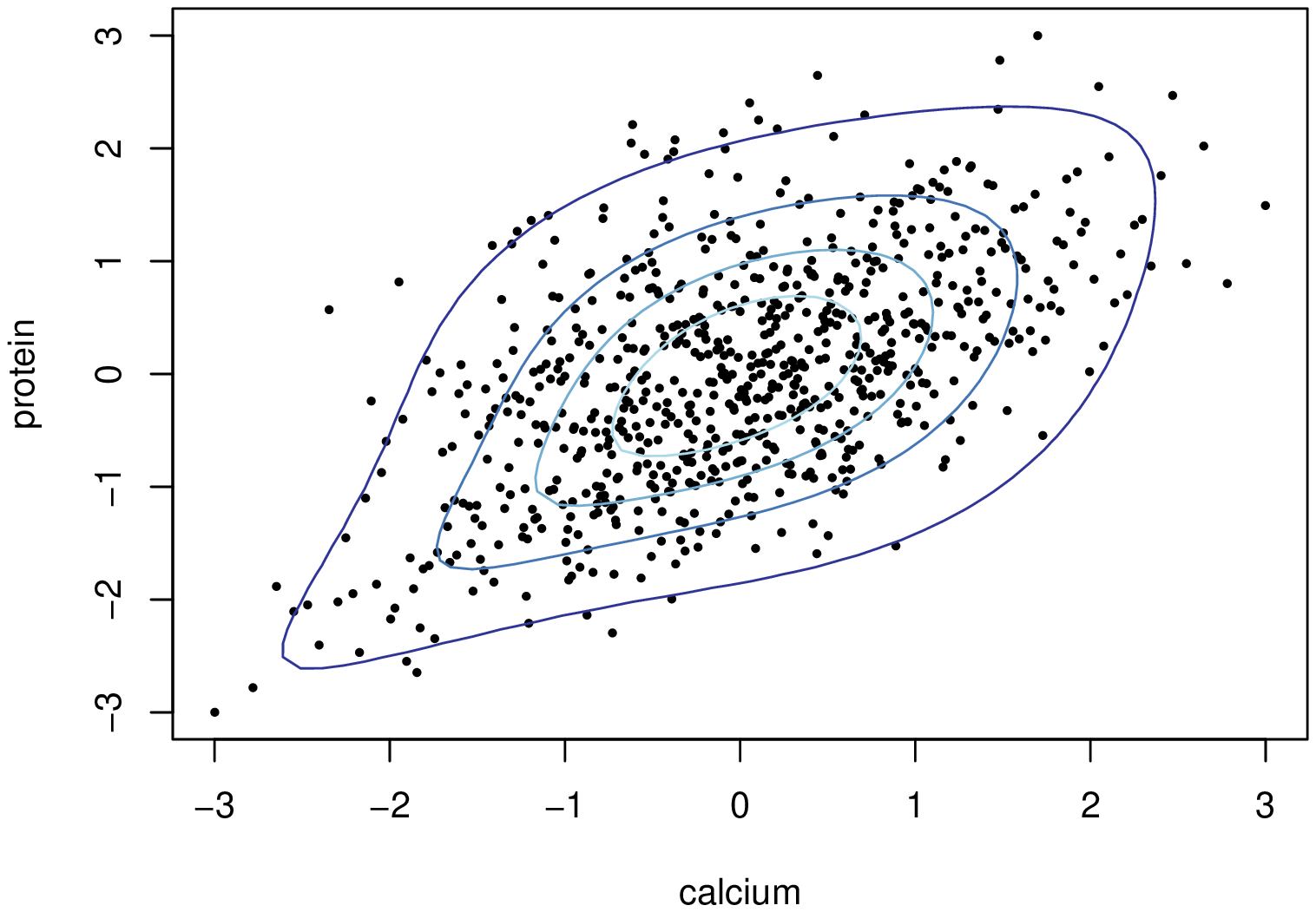}\\
\hline
\end{tabular}

\end{center}
\end{figure}

Then we exploit the use of  the $K$-FNM copula  
to construct a plausible copula family to represent the joint distribution of both pairs of variables. It has been revealed, that for both  pairs $K=2$ mixture components are sufficient to describe their dependence.   Table \ref{FNM-fit} gives the AICs and estimated 2-FNM copula parameters, along with their standard errors.  
The AICs show, that  between the intakes of calcium and iron and between the intakes of calcium and protein the 2-FNM copula provides   better fit than the survival BB7, since the AIC has been improved by  $3.1=-240.6-(-243.7)$ and $7.1=-284.6-(-291.7)$, respectively.  
In Figure \ref{app-contours} we depict the estimated contour plots of the 2-FNM and survival BB7 copulas  with standard normal margins, along with the bivariate normal scores plot for the pairs identified as asymmetric  in the nutrient data set.
From the plots, it  is revealed  that the 2-FNM copula provides a nearly identical or even better  representation of the joint distribution compared to the survival BB7 (best fit amongst the existing parametric families of copulas).

\section{\label{sec-discussion} Discussion}

We have proposed the $K$-FNM parametric family of bivariate copulas and  demonstrated that the new family  is so flexible, it removes the ad-hoc constraints on the tails of existing parametric copula families, and is able to handle various dependence patterns that appear in the existing parametric bivariate copula families.

There exist  many bivariate copula families, and as the new copula family can ``nearly" approximate any of these, the selection of the appropriate copula family among many candidates can be subsided by solely using the $K$-FNM copula. This applies when the data are continuous and have weak to moderate dependence and when the data are discrete for any different strength of dependence.

Given that bivariate copulas are building blocks for many multivariate dependence models such as the vine (e.g., \citealt{nikoloulopoulos&joe&li11,panagiotelis&czado&joe12,Dissmann-etal-2013-csda}) and factor (e.g., \citealt{krupskii-joe-2013,nikoloulopoulos&joe12,
Kadhem&Nikoloulopoulos-2021}) copula  models, there is much potential of the proposed copula for building up more complex multivariate dependence models.   Future research will focus on exploring this potential in modelling real multivariate  datasets that have  complex dependence structures.

\section*{Acknowledgements}
The simulations presented in this paper were carried out on the High Performance Computing Cluster supported by the Research and Specialist Computing Support service at the University of East Anglia.

\end{document}